\documentclass[12pt]{iopart}

\usepackage{iopams}  
\usepackage{setstack}

\usepackage{bbm}
\usepackage{bm}
\usepackage{amssymb}
\usepackage{amsthm}
\usepackage{graphicx}	
\usepackage{subfigure}
\usepackage{rotating}
\usepackage{mathrsfs}
\usepackage{wasysym}
\usepackage{stmaryrd}

\usepackage{tensor}

\newcommand{ \mv }[1]{ \mathbf{ #1 } }

\newcommand{\mA}{\mathbb{A}}
\newcommand{\mW}{\mathbb{W}}

\newcommand{ \sC }{ \mathcal{C} }

\newcommand{ \sD }{ \mathcal{D} }
\newcommand{ \sH }{ \mathcal{H} }

\newcommand{ \sN }{ \mathcal{N} }
\newcommand{ \sE }{ \mathbb{E} }

\newcommand{\nn}{\nonumber}

\begin{document}

\title[Controlling spontaneous-emission noise in a continuously-monitored BEC]{Controlling spontaneous-emission noise in measurement-based feedback cooling of a Bose-Einstein condensate}

\author{M R Hush$^1$, S S Szigeti$^2$, A R R Carvalho$^{3,4}$ and J J Hope$^{3}$}

\address{$^1$School of Physics and Astronomy, University of Nottingham, Nottingham, NG7 2RD, United Kingdom}
\address{$^2$ARC Centre for Engineered Quantum Systems, The University of Queensland, Brisbane, QLD 4072, Australia}
\address{$^3$Department of Quantum Science, Research School of Physics and Engineering, The Australian National University, Canberra, ACT 0200, Australia}
\address{$^4$ARC Centre for Quantum Computation and Communication Technology, The Australian National University, Canberra, ACT 0200, Australia}

\ead{michael.hush@nottingham.ac.uk}

\date{\today}

\begin{abstract}
Off-resonant optical imaging is a popular method for continuous monitoring of a Bose-Einstein condensate (BEC). However, the disturbance caused by scattered photons places a serious limitation on the lifetime of such continuously-monitored condensates. In this paper, we demonstrate that a new choice of feedback control can overcome the heating effects of the measurement backaction. In particular, we show that the measurement backaction caused by off-resonant optical imaging is a multi-mode quantum-field effect, as the entire heating process is not seen in single-particle or mean-field models of the system. Simulating such continuously-monitored systems is possible with the number-phase Wigner (NPW) particle filter, which currently gives both the highest precision and largest timescale simulations amongst competing methods. It is a hybrid between the leading techniques for simulating non-equilibrium dynamics in condensates and particle filters for simulating high-dimensional non-Gaussian filters in the field of engineering. The new control scheme will enable long-term continuous measurement and feedback on one of the leading platforms for precision measurement and the simulation of quantum fields, allowing for the possibility of single-shot experiments, adaptive measurements and robust state-preparation and manipulation.   

\end{abstract}


\vspace{2pc}

\submitto{\NJP}

\maketitle

\section{Introduction}
A Bose-Einstein condensate (BEC) in a dilute atomic gas can be well-isolated from its environment, and is highly controllable using a combination of optical, rf and magnetic fields \cite{Leggett:2001}.  This gives it the potential to address a broad range of research questions in fundamental and applied science, including studies in quantum non-equilibrium thermodynamics \cite{Kinoshita:2006,Barnett:2011}, entanglement of massive particles \cite{Kheruntsyan:2005} and the quantum simulation of both cosmological phenomena, such as Hawking radiation emitted from a black hole's event horizon \cite{Barcelo:2001}, and phase transitions in condensed matter, including superconductivity and quantum magnetism \cite{Buluta:2009}. Furthermore, Bose-condensed sources are likely to be key components in a range of future technologies, such as improved precision inertial sensors based on atom interferometry, where the sensitivities of current devices are limited by the properties of thermal sources \cite{Fixler:2007, Szigeti:2012, Robins:2013}.  This wealth of research opportunity has therefore led to great practical interest in the ability to control the spatial state of the BEC's quantum field. Research has predominantly focussed on open-loop control of the condensate by direct control of optical and magnetic potentials. In contrast, continuous measurement feedback control of BECs is rarely employed due to their perceived fragility under continuous measurement.  However, this intuition does not account for the advantages active feedback control can provide. Active feedback can control properties of the BEC using the information gained from the continuous measurement, which includes properties that would have been adversely affected by the backaction. Furthermore, active feedback provides robust and reliable behaviour that cannot be matched by open-loop control schemes \cite{Franklin:2009}. In particular, it has been shown that the linewidth of an outcoupled atom laser can be improved by continuously monitoring the BEC with off-resonant light and applying active feedback \cite{Wiseman:2001,Thomsen:2002,Wilson:2007,Szigeti:2009,Szigeti:2010,Szigeti:2013}. However, these prior analyses used single-particle or mean-field models of the system, and thus did not fully incorporate the multi-mode quantum-field effects of the measurement backaction. In this paper, we present the first analysis of a feedback-controlled BEC that \emph{does} include the crucial effects of measurement-induced \emph{spontaneous-emission noise}, by using the number-phase Wigner (NPW) particle filter to perform our simulations. Furthermore, we demonstrate that the effects of the spontaneous-emission noise can be cancelled with the application of an active feedback control that is tailored to the measurement being applied. This result demonstrates that there is no fundamental limit to the lifetime of a condensate due to off-resonant imaging, and also shows that improving the linewidth of outcoupled atom lasers from condensates is feasible with active feedback.

Continuous off-resonant measurement of BEC has been achieved in experiment \cite{Saba:2005,Andrews:1996,Stamper-Kurn:1999}, and potentially has numerous and varied applications. Recent proposals that rely upon the continuous off-resonant imaging of ultracold atomic systems include the generation of entanglement in hybrid optomechanical systems \cite{De-Chiara:2011}, the generation of coherence between two initially uncorrelated condensates \cite{Lee:2012}, an investigation into the emergence of classical dynamics from a quantum system \cite{Javanainen:2013}, and the simulation of condensed matter systems \cite{Romero-Isart:2012,Hauke:2013}. Unfortunately, in previous experiments the decoherence due to the spontaneous scattering of the photons placed hard limits on the lifetime of the condensate \cite{Andrews:1996,Higbie:2005}.  Such disruption could be mitigated by placing a BEC in a cavity \cite{Lye:2003,Hope:2004,Hope:2005}, but it is was previously unknown if this disruption could be cancelled entirely or if continuous monitoring places a fundamental limit on the lifetime of the BEC. Early work by Haine \emph{et al.} \cite{Haine:2004} showed that density disruptions in a condensate could be reduced using a feedback control based on simple adjustments to the BEC's confining potential, albeit in the unrealistic situation where measurement backaction was neglected. The effect of measurement backaction was incorporated in Wilson \emph{et al.} \cite{Wilson:2007}, which studied the effectiveness of simple damping feedback with a single-mode model of a BEC undergoing continuous centre-of-mass position measurement. Most relevant to this paper, previous analyses by Szigeti \emph{et al.} \cite{Szigeti:2009,Szigeti:2010}, who considered the semiclassical limit of a multi-mode quantum filtering model for an off-resonantly imaged, feedback-cooled BEC, suggested that active feedback could remove the disruption caused by off-resonant imaging entirely.  However, these models were analysed under the Hartree-Fock approximation, which assumes the condensate has a fixed number of atoms all in the same single-particle state, which in turn neglects part of the spontaneous-emission noise, making this analysis incomplete. Using the NPW particle filter to perform a multi-mode quantum-field analysis, we avoided such approximations and discovered that an \emph{additional} control mechanism is required.

The NPW particle filter allows for the simulation of a feedback controlled, continuously-monitored BEC that includes the quantum statistics ignored by the Hartree-Fock method. It has been developed over a sequence of papers by Hush \emph{et al.}, and is currently the most precise numerical method for simulating the quantum statistics of monitored BECs. The NPW particle filter method has two main components. The first is the Number-Phase-Wigner representation \cite{Hush:2010,Hush:2012,Hush:2012a} which falls into a class of stochastic methods based on phase-space representations, which includes both truncated Wigner (TW) \cite{Sinatra:2002,Blakie:2008,Steel:1998,Opanchuk:2013,Polkovnikov:2010}, positive-P (P$^{+}$) \cite{Gilchrist:1997,Deuar:2006,Steel:1998} and other variants \cite{Trimborn:2008, Korennoy:2011}. These methods have proven to be highly efficient at investigating the dynamics of BECs that cannot be probed by mean-field methods, including the generation of entanglement \cite{Kheruntsyan:2005}, squeezing \cite{Johnsson:2007} and thermal fluctuations \cite{Blakie:2008}. The second aspect of the NPW particle filter method is an application of an engineering technique called particle filters \cite{Smith:1993,Merwe:2000,Arulampalam:2002,Djuric:2003}.  Particle filters are the most efficient control theory methods available to continually estimate systems that are both high dimensional and nonlinear \cite{Arulampalam:2002}, such as a monitored BEC.

This paper aims to provide a complete investigation into the feedback controls that are required to remove the effects of spontaneous emission due to off-resonant imaging of a BEC.  In doing so, it also provides the first practical application of the NPW particle filter. Previous work has verified the accuracy of this simulation technique and compared its performance to other established methods \cite{Hush:2012}. To make this paper as accessible as possible, we provide an overview of the relevant numerical techniques, including the NPW particle filter, in section~\ref{prt:numerical}.   Section~\ref{prt:physics} investigates the heating induced by spontaneous-emission noise due to off-resonant optical imaging, and demonstrates how a novel control scheme can control this heating effect, leading to net cooling of the BEC to steady state. 

\section{Review of numerical techniques} \label{prt:numerical}

Here we give a brief overview of the numerical techniques used in this paper. For readers unfamiliar with classical stochastic methods \cite{Gardiner:2004,Jacobs:2010a}, quantum stochastic differential equations \cite{Gardiner:2004b, Wiseman:2010}, phase-space methods \cite{Gardiner:2004b, Walls:2008, Blakie:2008, Dowling:2007}, quantum measurement, filtering and control \cite{Ramon-van-Handel:2005, Bouten:2007, Wiseman:2010}, and BEC \cite{Dalfovo:1999, Leggett:2001,Pethick:2002}, the enclosed references are excellent introductions.   

In section~\ref{sec:condmasteqn} we introduce the conditional master equation for a BEC under continuous off-resonant imaging. In anticipation of future applications, we have kept the form of the conditional master equation as generic as possible. Section~\ref{sec:hartfockapp} introduces the Hartree-Fock approximation, which is the relevant semiclassical approximation for continuous off-resonant optical imaging of a BEC. Finally, in section~\ref{sec:npwparticlefitler} we give an overview of the NPW particle filter, including the algorithmic details for its implementation.

\subsection{Conditional master equation} \label{sec:condmasteqn}

We consider the simulation and analysis of a BEC undergoing off-resonant imaging and active feedback control. Continuously monitoring a BEC provides a signal that contains information regarding the density of the BEC. The measurement can be used not only to estimate the observable being measured, but also to give a best estimate (in the least-squares sense) for the full quantum state of the condensate. This can be achieved using a \emph{conditional master equation} \cite{Wiseman:1993,Wiseman:1994,Wiseman:2010}, which is better known as a \emph{filter} in the engineering community \cite{Belavkin:1983,Belavkin:1992,Bouten:2007}.  The conditional master equation of a BEC undergoing continuous off-resonant imaging is governed by the Stratonovich equation \cite{Wilson:2007,Dalvit:2002,Szigeti:2009,Szigeti:2010,Szigeti:2013,Hush:2012a}:
\begin{equation}
	\partial_t \hat{\rho} = -\frac{i}{\hbar}[\hat{H}(\mv{u}),\hat{\rho}] + \sum_i \sD[\hat{L}_i] \hat{\rho} + \sum_i \sC[\hat{L}_i] \hat{\rho} +  \sum_i \sH[\hat{L}_i] \hat{\rho} \; \eta_i(t),  \label{eqn:conBECmasts}
\end{equation}
where $\hat{\rho}$ is a density matrix encoding the full quantum field of the condensate conditioned on the measurement result. The first term describes the unitary dynamics of the system, which is determined by the Hamiltonian
\begin{equation}
	\hat{H}(\mv{u}) 	= \int d\mv{x} \; \hat{\psi}^\dag(\mv{x}) h(\mv{x},\mv{u}) \hat{\psi}(\mv{x}) + \frac{U}{2} \int d\textbf{x} \, \hat{\psi}^\dag(\mv{x})\hat{\psi}^\dag(\mv{x}) \hat{\psi}(\mv{x})\hat{\psi}(\mv{x}), \label{field_Ham}
\end{equation}
where $\hat{\psi}(\mv{x})$ is the field operator that annihilates an atom in the ground state at position $\mv{x}$ (we assume that the imaging light has been sufficiently detuned from the atomic resonance such that the excited state can be adiabatically eliminated). The field operators obey the commutation relations $[ \hat{\psi}(\mv{x}), \hat{\psi}^\dag(\mv{x}')] = \delta(\mv{x} - \mv{x}')$, where $\delta(\mv{x}-\mv{x}')$ is a Dirac delta function. The first term in equation~(\ref{field_Ham}) is the single-atom energy, governed by the single-particle Hamiltonian $h(\mv{x},\mv{u})$. Feedback control is included in this Hamiltonian via the vector of control signals, $\mv{u}(t)$, which in general depends upon the system state $\hat{\rho}$ at time $t$. The second term in equation~(\ref{field_Ham}) is the energy due to the two-body collisions between atoms, assuming a hard-sphere contact scattering potential of strength $U$. This is an excellent approximation for gases of ultracold alkali atoms \cite{Pitaevskii:2003}. 

The effect of the measurement is described by the remaining terms in the conditional master equation~(\ref{eqn:conBECmasts}). The superoperators 
\numparts
\begin{eqnarray}
	\mathcal{D}[\hat{c}]\hat{\rho} &\equiv \hat{c}\hat{\rho} \hat{c}^\dag -\frac{1}{2}\left(\hat{c}^\dag \hat{c}\hat{\rho}+\hat{\rho} \hat{c}^\dag \hat{c}\right), \label{eqn:sodeco} \\
	\mathcal{H}[\hat{c}] \hat{\rho} &\equiv \hat{c} \hat{\rho} + \hat{\rho} \hat{c}^\dag - \langle \hat{c}+\hat{c}^\dag \rangle \hat{\rho}, \label{eqn:somes} \\
	\mathcal{C}[\hat{c}] \hat{\rho} &\equiv \langle \hat{c} + \hat{c}^\dag \rangle \mathcal{H}[\hat{c}]\hat{\rho} -\frac{1}{2} \mathcal{H}[\hat{c}^2]\hat{\rho}+\langle \hat{c}^\dag \hat{c} \rangle \hat{\rho}-\hat{c} \hat{\rho} \hat{c}^\dag, \label{eqn:sostrat}  
\end{eqnarray}
\endnumparts
are the decoherence, innovations, and Stratonovich correction superoperators, respectively, for an arbitrary operator $\hat{c}$ where $\langle \cdot \rangle = \Tr[\cdot \hat{\rho}]$. The decoherence superoperator describes the effect of the measurement backaction on the system, while the innovations superoperator incorporates the information obtained from the measurement by continuously updating the density matrix. The specifics of the backaction and measurement signal depend upon the set of \emph{measurement operators}
\begin{equation}
	\hat{L}_i = \int d\mv{x} \; \hat{\psi}^\dag(\mv{x}) l_i(\mv{x}) \hat{\psi}(\mv{x}), \label{measurement_operators}
\end{equation}
where $l_i(\mv{x})$ are the density-moment generators, which depend upon the precise setup of the imaging optics. Note that since off-resonant imaging is a scattering process, the measurement operators are restricted to the form of equation~(\ref{measurement_operators}) to ensure that the particle number of the quantum gas is conserved. The stochastic nature of the random wavefunction collapse due to the set of measurement operators is included via the noises $\eta_i(t)$ (more on these shortly). The validity of the conditional master equation (\ref{eqn:conBECmasts}) is corroborated by previous theoretical examinations into the off-resonant imaging of BEC by Dalvit \emph{et al.} \cite{Dalvit:2002}, Szigeti \emph{et al.} \cite{Szigeti:2009,Szigeti:2010}, and others \cite{Thomsen:2002, Ruostekoski:1997,Corney:1998,Li:1998,Leonhardt:1999}. 

Physically, the conditional master equation (\ref{eqn:conBECmasts}) gives an estimate of the condensate as a function of the measurement signal(s). In principle, this conditional master equation could be integrated during an experiment to give a real-time estimate of the current quantum state of the BEC. Furthermore, this information could then be used to apply the required controls to the condensate, which would complete an active feedback loop. In this case, $\eta_i(t)$ corresponds to the difference between each measurement signal $Y_i(t)$ and the current estimate of the observable being measured \cite{Ramon-van-Handel:2005}. Explicitly, if the measurement signal changes by $\Delta Y_i$ within a time interval $\Delta t$, then
\begin{equation}
	\Delta \eta_i = \Delta Y_i -  \Tr[ (\hat{L}_i + \hat{L}_i^\dag) \hat{\rho}] \Delta t.
\end{equation} 
However, in this paper we only \emph{simulate} equation~(\ref{eqn:conBECmasts}). Fortuitously, this can be achieved by using equation~(\ref{eqn:conBECmasts}) and simply treating $\eta_i(t)$ as a Stratonovich stochastic integral \cite{Wong:1965,Gardiner:2004b,Bouten:2007,Wiseman:2010}. More specifically, we sample $\Delta \eta_i(t)$ from a Gaussian distribution with variance $1/\Delta t$, and then numerically integrate equation~(\ref{eqn:conBECmasts}) using an algorithm appropriate for Stratonovich stochastic differential equations (SDEs) (for some examples, see \cite{Kloeden:1999}). Each stochastic path of $\eta_i(t)$ now corresponds to a virtual measurement record corresponding to a single run of an `experiment'.

In practice, we simulate equation~(\ref{eqn:conBECmasts}) many times (each with a different realization of the stochastic noises, $\eta_i^{(p)}(t)$, resulting in different density matrices, $\hat{\rho}^{(p)}$, here indexed by $p$) and then average these \emph{paths} (or \emph{trajectories}) to get a result. We use the notation $\mA[ \cdot]$ to mean this procedure. For example, in order to find the average density over time, we simulate equation~(\ref{eqn:conBECmasts}) $P$ times and then compute 
\begin{equation}
	\mA\left[\langle \hat{\psi}^\dag(\mv{x}) \hat{\psi}(\mv{x}) \rangle \right] = \frac{1}{P}\sum_{p=1}^P \Tr\left[ \hat{\psi}^\dag(\mv{x}) \hat{\psi}(\mv{x}) \hat{\rho}^{(p)}\right],
\end{equation} 
These \emph{ensemble averages} describe the typical behaviour of the system. They are generally more useful to examine than individual stochastic paths, as conclusions based on individual paths may be spurious due to the stochastic nature of the paths. However, we emphasize that every integration of equation~(\ref{eqn:conBECmasts}) is physically meaningful, and thus each $\hat{\rho}^{(p)}$ can be thought of as a new `experiment'. We note that if there was no feedback then $\mA[\hat{\rho}]=\hat{\varrho}$ in the limit as $P \rightarrow \infty$, where $\hat{\varrho}$ is the density matrix governed by the \emph{unconditional} version of equation~(\ref{eqn:conBECmasts}).  \emph{This is not the case when feedback is active}, since feedback can produce non-Markovian behaviour that cannot be modelled by an unconditional (or traditional) master equation.

Unfortunately, simulating equation~(\ref{eqn:conBECmasts}) directly and exactly is impractical for even a modest number of particles. Suppose we limit the number of modes we simulate to $M$, and then truncate the number of excitations in each mode to $J$. Then the field operator can be written as $\hat{\psi}(\mv{x}) = \sum_{m=1}^M u_m(\mv{x}) \hat{a}_m$, where $\left\{u_m(\mv{x})\right\}_m$ are a set of orthonormal functions, and each mode $\hat{a}_m$ has a set of $J$ energy eigenstates labeled $|j_m\rangle$. The density matrix is then
\begin{equation}
	\hat{\rho} = \sum^{J}_{ {j_1,\cdots j_M = \, 0} \atop {j_1',\cdots j_M' =\, 0} } c_{j_1, \cdots j_M,j_1', \cdots j_M'} |j_1, \cdots j_M \rangle \langle  j_1', \cdots j_M'|.
\end{equation}
The memory required to store this density matrix is $J^{2M}$ times the memory required to store each complex number.  This exponential scaling of the size of the quantum field with the number of modes is an explicit example of the \emph{curse of dimensionality} that generally restricts the direct simulation of large quantum systems. Both the Hartree-Fock method and NPW particle filter are approximate techniques that circumvent this problem. However, as shown shortly, simulations of equation~(\ref{eqn:conBECmasts}) with the NPW particle filter retain more of the important quantum statistics than simulations using the semiclassical Hartree-Fock approximation.

\subsection{Hartree-Fock approximation} \label{sec:hartfockapp}

In many BEC experiments, the dynamics of the condensate depend minimally on correlations in the atomic field, and can thus be well-described by a mean-field theory. This is analogous to the case in classical optics, where the quantum fluctuations in the electric field can be ignored. There are a number of different approaches to constructing a semiclassical theory of BEC dynamics, which in most situations give identical equations of motion. Common to all approaches is the assumption that all the atoms of the condensate occupy the same single-particle quantum state. This allows the system to be modelled as a single macroscopic wavefunction (or order parameter) with a size \emph{independent} of the total number of particles, thereby circumventing the curse of dimensionality associated with the full quantum field. 

The semiclassical theory used in this paper is based on the Hartree-Fock approximation. This assumes that (a) the BEC is always in a state of fixed total number $N$, and (b) that all $N$ atoms occupy the same mode. This restricts the conditional density matrix in equation~(\ref{eqn:conBECmasts}) to the form
\begin{equation}
	\hat{\rho}_{\text{HF}}(t) =\bigotimes_{i=1}^N \left(\int d\mv{x} \, d\mv{y} \, \phi^*(\mv{x},t)\phi(\mv{y},t) |\mv{x} \tensor*{\rangle}{_i} \tensor*[_i]{\langle}{}\mv{y} |\right), \label{eqn:denssamewave}
\end{equation}
where $\hat{\rho}_{\text{HF}}$ is written in first quantized form with $|\mv{x} \rangle_i$ the position eigenstate of the $i$th atom and $\phi(\mv{x},t)$ the position-basis wavefunction. This approximation fixes all the quantum statistics of the condensate; for example 
\begin{equation}
	\langle \hat{\psi}^\dag(\mv{x}) \hat{\psi}(\mv{y}) \rangle = N \phi^*(\mv{x},t)\phi(\mv{y},t) \label{eqn:twobodydenhart} \\
\end{equation}
and
\begin{equation}
	\langle (\hat{\psi}^\dag(\mv{x}))^2 (\hat{\psi}(\mv{x}))^2 \rangle = N(N-1) |\phi(\mv{x},t)|^4. \label{eqn:twobodycorhart}
\end{equation}
Thus, we cannot describe squeezed states, thermal states or any other exotic quantum states with the Hartree-Fock approximation. As described in Szigeti \emph{et al.} \cite{Szigeti:2010}, equations~(\ref{eqn:denssamewave}) and (\ref{eqn:twobodydenhart}) can be used to find the equation of motion for the one-body correlation matrix~(\ref{eqn:twobodydenhart}), which then gives the following equation of motion for the \emph{unnormalized} wavefunction:
\begin{eqnarray}
	\partial_t \tilde\phi(\mv{x},t) 	&=  \Big\{ -\frac{i}{\hbar}\left(h(\mv{x},\mv{u}) + (N-1)\frac{U}{n_\phi(t)} |\tilde{\phi}(\mv{x},t)|^2\right) \nn \\
 							&+ \sum_i \left(2 l_i(\mv{x}) L_i^\phi(t)-l_i(\mv{x})^2 + l_i(\mv{x}) \; \eta_i(t)\right) \Big\} \tilde{\phi}(\mv{x},t), \label{eqn:monBEChartreefock}
\end{eqnarray}
where $\tilde{\phi}(\mv{x},t)$ is related to the normalized wavefunction by 
\begin{equation}
	\phi(\mv{x},t) = \frac{\tilde{\phi}(\mv{x},t)}{\sqrt{n_\phi(t)}},
\end{equation}
and we have defined
\begin{eqnarray}
	n_\phi(t) 	&\equiv\int d\mv{x}\, |\tilde{\phi}(\mv{x},t)|^2, \\
	L_i^\phi(t) &\equiv \int d\mv{x}\, l_i(\mv{x}) \frac{|\tilde{\phi}(\mv{x},t)|^2}{n_\phi(t)}. 
\end{eqnarray}
Numerical integration of the unnormalized evolution equation is preferred as it is much faster and possesses increased numerical stability over the normalized equation of motion. 
Equation~(\ref{eqn:monBEChartreefock}) must be integrated for each virtual measurement record $\eta_i(t)$, although this does not need to be done simultaneously. Each integration corresponds to a path or an `experiment'. We note that this equation reduces exactly to the equation of motion for a single particle when $N=1$. 

The Hartree-Fock approximation is only valid for condensates well below the phase transition temperature. Even if this condition is satisfied for the  initial state, it may cease to be true due to the heating caused by the off-resonant imaging.  However, we will see that for measurements that probe the condensate weakly, the Hartree-Fock approximation remains valid. 

It might be wondered what advantages the Hartree-Fock method has over the coherent-state-based mean-field approximation that is traditionally used to derive the Gross-Pitaevskii equation (GPE), given that both approximations result in simulations of identical dimensionality.  The answer is that applying a coherent-state-based mean-field approximation to equation~(\ref{eqn:conBECmasts}) produces a GPE-like equation that is numerically unstable \cite{Szigeti:2010}. This is thought to be due to the effect of the measurement. Off-resonant imaging rapidly gathers information about the total number of atoms in the condensate, projecting the system into a subspace with a well-known total atom number. The Hartree-Fock approximation has a fixed total number of atoms and continues to be valid during this measurement process. In contrast, the mean-field approximation does not have a fixed total number of atoms, and thus gives an inaccurate description of the system. Note that in the case when a BEC is not being monitored, the Hartree-Fock and mean-field approximations are virtually identical. For if we set $l_i = 0$ and assume that $N \gg 1$, then equation~(\ref{eqn:monBEChartreefock}) reduces to the GPE. Despite these important differences, both the Hartree-Fock approximation and coherent-state-based mean-field approximation are in the same class of semiclassical approximations where the quantum statistics are fixed. We now examine a simulation method that does not fix the quantum statistics, and is therefore able to simulate a much larger class of physical phenomena.

\subsection{NPW particle filter} \label{sec:npwparticlefitler}

The NPW particle filter is currently the most precise method for simulating the long timescale dynamics of a continuously-monitored BEC without fixing the quantum statistics \cite{Hush:2010}.  Instead of integrating a single complex function, as required by the Hartree-Fock approximation, the NPW particle filter requires the simultaneous integration of multiple complex functions, each with its own real-valued \emph{weight} and additional stochastic noise. The weights determine which one of the complex functions best match the current measurement record, while the noise acts as a correction for the backaction due to the measurement. Observables are then calculated by using a weighted average over these complex functions. 

The approximations used in the derivation of the NPW particle filter are detailed in \ref{apx:derivNPWparfil}.  Like the truncated Wigner (TW) phase-space method \cite{Sinatra:2002,Blakie:2008,Steel:1998,Opanchuk:2013,Polkovnikov:2010}, it is valid when the number of atoms $N$ is much larger than the number of modes $M$ being simulated. For coherent evolution, the NPW particle filter and TW phase-space method produce essentially identical simulations \cite{Hush:2012}. TW can also be used to simulate open quantum systems, an early example being \cite{Drummond:1993}. However, the NPW particle filter is more numerically stable and accurate than TW when a system is continuously-monitored. This is because the TW representation generates non-semi-positive definite diffusion when applied to equation~(\ref{eqn:conBECmasts}), which makes it numerically unstable \cite{Hush:2012,Hush:2012a}. This might come as a surprise, since TW is guaranteed to produce semi-positive definite diffusion when applied to \emph{deterministic} master equations. However, the assumptions required for this proof fail to hold in the case of \emph{stochastic} master equations. In contrast, the NPW particle filter has no such issue. We can think of the NPW particle filter as being an additional member of the so-called c-field techniques for simulating the non-equilibrium dynamics of BECs \cite{Blakie:2008, Gardiner:2002}, and one which is of most use when the condensate is being monitored. 
 
Phase-space methods typically require multiple integrations of stochastic equations to produce quantum averages. However, the addition of monitoring to the problem requires a set of \emph{weighted} stochastic differential equations (WSDE) that are simulated in parallel.  The concept of a weighting and weighted averages is common in statistics and meta-studies \cite{Cochran:1937,Meier:1953,Cochran:1954}. Concepts similar to weights have also been used in phase-space methods \cite{Deuar:2001,Deuar:2002}, quantum measurement theory \cite{Wiseman:1996} and quantum control \cite{Gambetta:2005}. Of the many applications, the techniques that are closest in spirit to our approach are so-called particle filters \cite{Smith:1993,Merwe:2000,Arulampalam:2002,Djuric:2003}, which have become a leading technology in the field of engineering for classical filtering problems where Gaussian approximations do not apply. Examples include image tracking \cite{Ristic:2004,Khan:2004,Okuma:2004}, GPS navigation \cite{Gustafsson:2002} and financial modelling \cite{Omori:2007,Djuric:2012}. Particle-filter-like simulations appear to have first been applied to quantum systems in the context of simulating monitored BECs \cite{Hush:2009}. However, very soon after this publication they were also applied by Jacobs \cite{Jacobs:2010} in the simulation of conditional master equations with imperfect detection efficiency. More generally, the emergence of quantum control as an important topic in both theory and experiment is seeing a growth in the application of advanced techniques from engineering, such as particle filters, to quantum mechanical problems.   

The NPW particle filter corresponding to equation~(\ref{eqn:conBECmasts}) is derived by mapping this conditional master equation to the NPW representation \cite{Hush:2010}, applying a series of approximations valid in the limit where $N\gg M$ (the details can be found in \cite{Hush:2012}), and finally transforming this representation into the NPW particle filter \cite{Hush:2009}. An overview of this process is presented in \ref{apx:derivNPWparfil} and in more detail in \cite{Hush:2012a}. The final result is a \emph{swarm} of $K$ complex functions $\alpha^{(k)}(x)$ with corresponding weights $w^{(k)}$, indexed with $k \in (1,K)$, that obey the following differential equations\footnote{For notational compactness, we suppress the time index for the functions $\alpha^{(k)}$ and weights $w^{(k)}$.}:  
\numparts
\label{eqn:conBECnpwsdes}
\begin{eqnarray}
\partial_t \alpha^{(k)}(\mv{x}) = -i \Big(\frac{1}{\hbar}(h(\mv{x},\mv{u}) + U|\alpha^{(k)}(\mv{x})|^2) + \sum_i l_i(\mv{x}) \zeta_i^{(k)}(t) \Big) \alpha^{(k)}(\mv{x}),  \\
\partial_t w^{(k)} =  2  \sum_i \Big( 2 L_i^{(k)} \mW[L_i^{(\cdot)}] - (L_i^{(k)})^2 + 2 L_i^{(k)} \eta_i(t) \Big)w^{(k)},  \label{eom:weights}
\end{eqnarray}
\endnumparts
where $\zeta_i^{(k)}(t)$ is a `fictitious' Stratonovich noise increment, which is generated independently for each of the $K$ paths, 
\begin{equation}
	L_i^{(k)} = \int d\mv{x} \; l_i(\mv{x}) \left(|\alpha^{(k)}(\mv{x})|^2 - \frac{1}{2}\delta(\mv{x})\right),
\end{equation}
 and 
\begin{equation}
\mW[f(\alpha^{(\cdot)}(\mv{x}))] \equiv  \frac{ \sum_{k=1}^K w^{(k)} f(\alpha^{(k)}(\mv{x}))}{\sum_{k=1}^K w^{(k)}} \label{eqn:weigtedavg}
\end{equation}
is the definition of the weighted average over the swarm. The Stratonovich noises $\eta_i(t)$ are virtual measurement records (corresponding to separate repetitions of the `experiment'), and are the same for all $k$. Importantly, no information is exchanged between different `experiments', so each swarm can be integrated independently. This is not strictly true for each element of the swarm. However, the different $k$ are only `weakly' coupled via the weighted averages.

Expectations of observables are calculated by using the weighted average defined in equation~(\ref{eqn:weigtedavg}). As specified in \cite{Hush:2010}, each moment generated is related to a specific ordering of operators. In this paper, we only consider observables that can be generated from the one-body correlation matrix, calculated using:
\begin{equation}
\langle \hat{\psi}^\dag(\mv{x}) \hat{\psi}(\mv{y}) \rangle = \mW[\alpha^*(\mv{x})\alpha(\mv{y})] - \frac{1}{2} \delta(\mv{x}-\mv{y}).\label{eqn:expectationValues}
\end{equation}
For comparison with equation~(\ref{eqn:twobodycorhart}), the expectation of the local number squared is given by
\begin{equation}
	\langle (\hat{\psi}^\dag(\mv{x}))^2 (\hat{\psi}(\mv{x}))^2 \rangle = \mW[|\alpha(\mv{x})|^4] - 2\mW[|\alpha(\mv{x})|^2] \delta(0) + \frac{1}{2} \delta(0)^2. \label{eqn:twobodycorNPW}
\end{equation}
In practice, the presence of the infinite quantity $\delta(0)$ causes no difficulty. For if the field $\alpha(\mv{x})$ is approximated as a discretized grid of points (which is required for numerical simulation), then $\delta(0) \approx 1/\Delta x$, where $\Delta x$ is the spacing between points on the position-space grid. This feature is common to all phase-space simulations of quantum fields \cite{Polkovnikov:2010}. Unlike the Hartree-Fock method, expectations computed via equation~(\ref{eqn:twobodycorNPW}) require weighted averages over the swarm. Hence, higher-order moments generated by the NPW particle filter are not locked to a mean field and can fluctuate dynamically. This freedom is what allows the NPW particle filter, and more generally phase-space methods, to investigate the dynamic quantum statistics of a system.

Simulating equation~(\ref{eqn:conBECmasts}) via the NPW particle filter~(15) requires appropriate initial conditions for the swarm of complex functions and the weights. Much like the TW method, these initial conditions are sampled from a quasi-probability distribution, and so in general are \emph{not} deterministic. \ref{apx:sampNPWparfil} gives the quasi-probability distribution from which to sample the initial conditions for a BEC, and further outlines an algorithm for this initial sampling. Analogous to the Hartree-Fock method, each sample has a quantized and fixed number of particles.

The evolution of the weights [equation~(\ref{eom:weights})] typically generates an exponential divergence between the largest and smallest weights. This can lead to poor sampling and thus an inaccurate estimate of the current quantum state of the BEC. There are a few numerical tricks to compensate for this issue. Firstly, it is more efficient to integrate the weights in log space, meaning $w_l =\ln(w)$ is integrated instead of directly integrating $w$. Secondly, it is important to frequently renormalize the weights to prevent them from becoming too small or too large. This is achieved by first calculating the current norm: 
\begin{equation}
	\bar{\omega} = \sum_{k=1}^K \omega^{(k)}, 
\end{equation}
and then reassigning weight values as follows:
\begin{equation}
	\frac{\omega^{(k)}}{\bar{\omega}} \rightarrow \omega^{(k)},
\end{equation} 
where we use $\rightarrow$ to mean `write to' in an algorithmic sense. Immediately after reassigning weights, $\bar{\omega} = 1$. The final and most important numerical consideration is \emph{resampling}.  As the maximum weight $\omega_{\text{max}} = \max_k \omega^{(k)}$ becomes exponentially larger than the smallest weight $\omega_{\text{min}} = \min_k \omega^{(k)}$, the complex field with the smallest weight becomes negligible. This causes poorer sampling, and furthermore, the memory being used to store this complex field is wasted.  We can therefore neglect the fields which have very small weights (compared to $\omega_{\text{max}}$) and replace them with an appropriately rebalanced sample from the maximum weighted complex field. A simple algorithm for this process is presented below. We assume the user sets a tolerance $\epsilon_{\text{tol}}$ such that if the ratio of a weight to the largest weight is below $\epsilon_{\text{tol}}$, then that weight is neglected.
\begin{enumerate}
	\item  \label{lst:breedstart} Find the index of the largest weight, $k_{\omega_{\text{max}}}$, and store the weight's value: 
	\begin{equation*}
		\omega^{(k_{\omega_{\text{max}}})} = \max_k w^{(k)} \rightarrow w_{\text{max}}.
	\end{equation*}
	\item Find the index of the smallest weight, $k_{\omega_{\text{min}}}$, and store the weight's value: 
	\begin{equation*}  
		\omega^{(k_{\omega_{\text{min}}})} = \min_k w^{(k)} \rightarrow w_{\text{min}}.
	\end{equation*}
	\item If the minimum weight is relatively smaller than our tolerance, $w_{\text{min}} / w_{\text{max}} < \epsilon_{\text{tol}}$, continue to (\ref{lst:breedcont}), otherwise skip to (\ref{lst:breedend}).
	\item \label{lst:breedcont} Overwrite the amplitude possessing the minimum weight with the amplitude whose corresponding weight is the maximum weight: 
	\begin{equation*}
		\alpha^{(k_{\omega_{\text{max}}})}(\mv{x}) \rightarrow \alpha^{(k_{\omega_{\text{min}}})}(\mv{x}).
	\end{equation*}
	\item Rebalance the weights: $w_{\text{max}}/2 \rightarrow w^{(k_{\omega_{\text{min}}})}$ and $w_{\text{max}}/2 \rightarrow w^{(k_{\omega_{\text{max}}})}$.
	\item Go back to (\ref{lst:breedstart}), repeat steps until all negligible weights have been removed.
	\item Finish and return to the integration of equations~(15). \label{lst:breedend}
\end{enumerate}
This algorithm, with a few minor enhancements, was used in the simulations presented in this paper. Note that this algorithm is \emph{not} the optimal solution to the resampling problem. Considerably more advanced and efficient resampling techniques have been developed, and can be found within the particle filtering literature. A tutorial by Arulampalam \emph{et al.} \cite{Arulampalam:2002} gives an excellent introduction to these resampling techniques. 

The numerical techniques required to simulate equation~(\ref{eqn:conBECmasts}) using either the Hartree-Fock approximation or the NPW particle filter have been presented. We now apply and compare both of these methods to a condensate undergoing off-resonant imaging and active feedback damping. This comparison, and the use of the NPW particle filter to design an effective quantum-noise feedback control, are the key results of this paper.

\section{Cancelling spontaneous-emission noise} \label{prt:physics}
\begin{figure}[h!tb]
\includegraphics[scale=0.45 ]{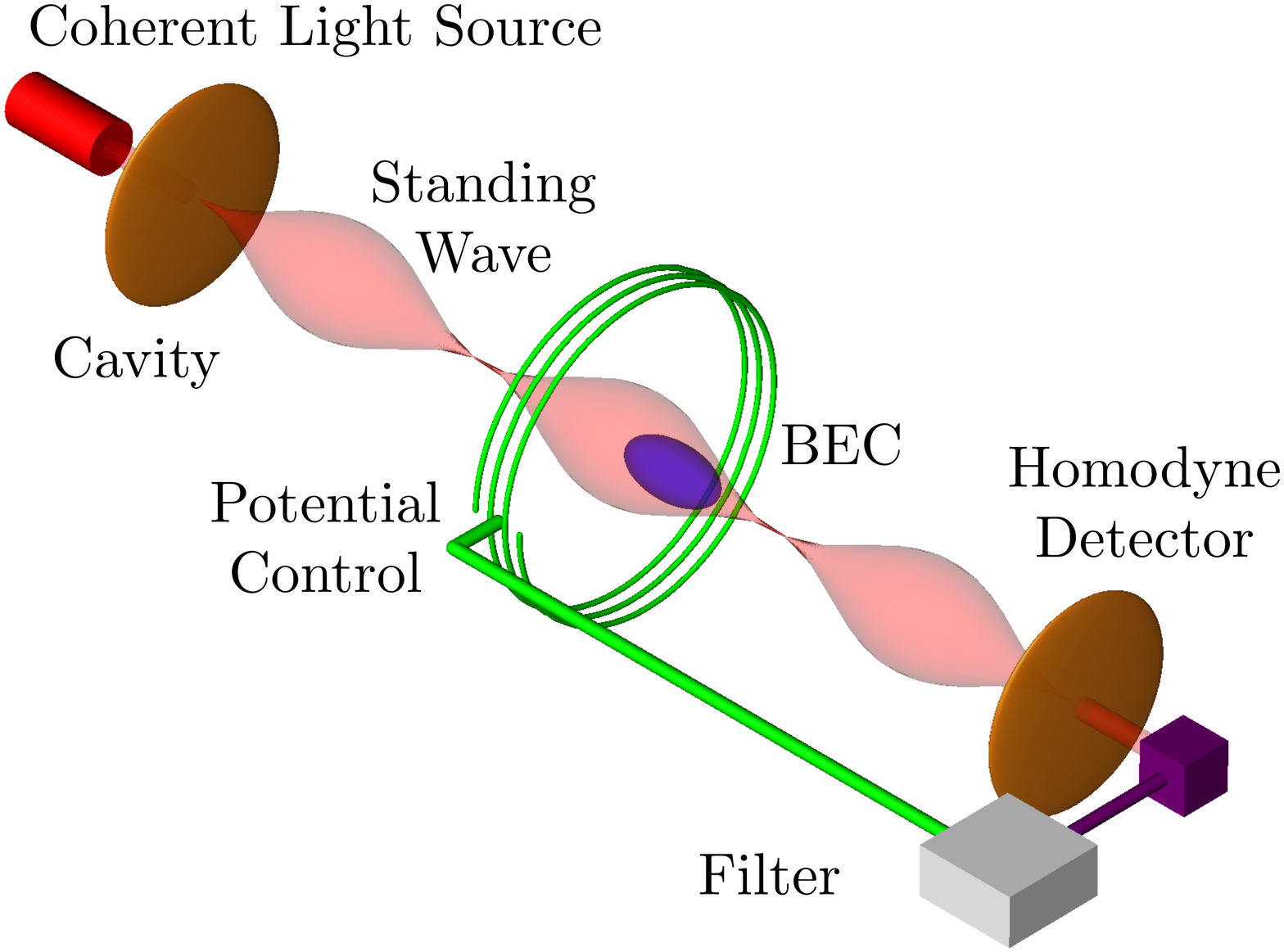}
\caption{Schematic diagram of a BEC under feedback control using a continuous cavity-mediated measurement. The condensate is coupled to the mode of a cavity (depicted by the red cosine-squared surface). Then a homodyne measurement of light output from the cavity gives a measurement signal proportional to a density moment of the atomic cloud. The green coils represent controls that modify the trapping field with the aim of driving the BEC to a low energy steady state, the trapping field itself is not shown.}
\label{fig:cavBECdiag}
\end{figure}

We consider the continuous monitoring of an harmonically trapped BEC under two common off-resonant imaging systems: a cavity-mediated measurement and phase-contrast imaging, both shown schematically in figure~\ref{fig:cavBECdiag} and figure~\ref{fig:directBECdiag}. Our aim is to determine whether active feedback control can be used to reduce the excitations in the condensate density caused by the measurement backaction (or, indeed, excitations due to any other process). In particular, we examine the effectiveness of a linear control that alters the trapping position \cite{Haine:2004,Szigeti:2009, Szigeti:2010,Szigeti:2012} using both the semiclassical Hartree-Fock approximation and the NPW particle filter. We show that in both imaging apparatuses there exist experimentally accessible regimes where the measurement induces additional heating in the condensate that \emph{cannot} be removed by simple feedback to the trapping position. This additional heating is not modelled by the Hartree-Fock simulations, and thus is a quantum multi-mode effect which we call \emph{spontaneous-emission noise} (for reasons that become clear shortly). Fortunately, as we also show in section~\ref{sec_q_noise_control}, it is possible to design a control that cancels the spontaneous-emission noise, allowing the continuously-monitored BEC to be cooled to a steady state. 

\subsection{Cavity-mediated measurement}
We first consider active feedback damping of a BEC of two-level atoms undergoing continuous cavity-mediated measurement, as shown in figure~\ref{fig:cavBECdiag}. Damping via feedback to the trapping position has been studied extensively for neutral atoms and optomechanical resonators \cite{Doherty:1999,Doherty:2000,Doherty:2012, Szigeti:2013}, and the measurement has been demonstrated experimentally for optomechanical resonators \cite{Szorkovszky:2011} and ultracold atoms \cite{Purdy:2010,Murch:2008b}.  Light is passed through an optical cavity containing the trapped BEC. We consider the regime of `good measurement', where the light is far-detuned from the atomic resonance and the cavity mirrors are lossy (i.e. large cavity linewidth), thereby allowing the atomic excited state and cavity dynamics to be adiabatically eliminated, respectively \cite{Doherty:1999}. Then the phase of the transmitted light contains a signal proportional to the overlap of the atomic density with the spatial envelope of the optical field in the cavity.  A homodyne measurement of this phase can be used to continually update the estimate of the atomic state. We model this estimate of the system, conditioned on the measurement record, using a conditional master equation (also called a quantum filtering equation). Energy can be removed from the BEC by continuously adjusting the mean position of the trap by an amount proportional to the negative of the bulk momentum of the BEC (which can be estimated from the quantum filter). The equation of motion for the filter that governs the single-atom version of this system is found in \cite{Doherty:1999,Gardiner:2004b}. This single-atom conditional master equation is a good model for a weakly-interacting BEC undergoing a continuous cavity-mediated measurement provided a collective mode of the atomic-field can be coupled to a single mode of the cavity \cite{Murch:2008}. Specifically, the Stratonovich quantum filtering equation for the conditional state of the atomic field is:
\begin{equation}
	\partial_t \hat{\rho} 	= -i[\hat{H}_F(u_s),\hat{\rho}] + \gamma \, \mathcal{D}[\hat{C}_\xi] \hat{\rho} + \gamma \, \mathcal{C}[\hat{C}_\xi] \hat{\rho}  + \sqrt{\gamma} \mathcal{H}[\hat{C}_\xi] \hat{\rho} \; \eta(t).  \label{eqn:cavBECmasts}
\end{equation}
For notational convenience, we have written energy in units of $\hbar \omega$, position in units of $\sqrt{\hbar/(m\omega)}$, and time in units of $1/\omega$, where $\omega$ is the frequency of the harmonic trapping potential and $m$ is the mass of an individual atom. The unitary dynamics are governed by the atomic-field Hamiltonian
\begin{equation}
	\hat{H}_F = \int dx \; \hat{\psi}^\dag(x) h_F(x,u_s) \hat{\psi}(x),
\end{equation}
where $\hat{\psi}(x)$ is the one-dimensional field operator satisfying $[\hat{\psi}(x), \hat{\psi}^\dag(x')] = \delta(x - x')$, and the single-particle Hamiltonian
\begin{equation}
	h_F(x,u_s) = \frac{1}{2}\left(-\partial_x^2 + x^2\right) + u_s \frac{\langle \hat{P} \rangle}{ \langle \hat{N} \rangle}x
\end{equation}
contains the kinetic energy, potential energy due to the harmonic trap, and the linear feedback of strength $u_s > 0$. $\hat{P}$ and $\hat{N}$ are the many-body momentum and number operators, respectively, defined as
\begin{eqnarray}
	\hat{P} &= \int dx\; \hat{\psi}^\dag(x) (-i \partial_x) \hat{\psi}(x), \\
	\hat{N} &= \int dx \; \hat{\psi}^\dag(x) \hat{\psi}(x).
\end{eqnarray}
The strength of the measurement is given by $\gamma$, which is proportional to the effective atom-cavity coupling rate and inversely proportional to the cavity linewidth.  Increasing $\gamma$ gathers more information about the system per unit time, however it also increases the rate of heating due to the measurement backaction. The observable that is measured is given by the measurement operator
\begin{equation}
	\hat{C}_\xi = \int dx \; \hat{\psi}^\dag(x) c_{\xi}(x) \hat{\psi}(x), 
\end{equation}
where $c_{\xi}(x) = \cos^2(\xi x - \pi/4)$ is proportional to the intensity of the intracavity optical field. The dimensionless length $\xi = 2 \pi x_{\text{HO}}/\lambda$ is the ratio of the natural length scale of the trap, $x_{\text{HO}} = \sqrt{\hbar/(m\omega)}$, and the wavelength $\lambda$ of the optical cavity \cite{Doherty:1999}. 

In the limit $\xi \ll 1$, $c_\xi(x) \approx 1/2 + \xi x$, and so the cavity-mediated measurement is approximately a measurement of the centre-of-mass position of the condensate. A BEC undergoing a continuous position measurement was used as a testing ground for the NPW simulation method due to the existence of reliable benchmarks \cite{Hush:2012}. However, in contemporary experiments a very small $\xi$ can be difficult to achieve, and so we consider the more moderate values of $\xi=0.1$ and $\xi=0.5$. 

We integrated equation~(\ref{eqn:cavBECmasts}) using both the restrictive Hartree-Fock approximation and the more complete NPW particle filter. Since equation~(\ref{eqn:cavBECmasts}) is a special case of equation~(\ref{eqn:conBECmasts}), we can use the general result of equation~(\ref{eqn:monBEChartreefock}) to find the equation of motion for the unnormalized macroscopic `wavefunction' $\tilde{\phi}(x,t)$:
\begin{equation}
	\fl \partial_t \tilde{\phi}(x,t) = \Big\{-i h_F(x, u_s) + \gamma \left(2 c_\xi(x) C_\xi^\phi(t) -c_\xi(x)^2 \right) + \sqrt{\gamma} \; c_\xi(x) \; \eta(t) \Big\} \tilde{\phi}(x,t), \label{eqn:cavBEChart}
\end{equation}
where 
\begin{eqnarray}
	C_\xi^\phi(t) &= \int dx\; c_{\xi}(x) \frac{|\tilde{\phi}(x,t)|^2}{n_\phi(t)}, \\
	n_\phi(t)	&= \int dx\; |\tilde{\phi}(x,t)|^2.
\end{eqnarray}
All the atoms have the same wavefunction $\tilde{\phi}(x)$ under the Hartree-Fock approximation. Similarly, the NPW particle filter for equation~(\ref{eqn:cavBECmasts}) is given by the general result of equations~(15):
\numparts
\label{eqn:cavBECnpwsdes}
\begin{eqnarray}
\partial_t \alpha^{(k)}(x) 	&= -i\Big(h_F(x,u_s) + \sqrt{\gamma} c_\xi(x) \; \zeta^{(k)}(t)\Big) \alpha^{(k)}(x), \\
\partial_t w^{(k)} 		&=  2 \left\{ \gamma \left(2C_\xi^{(k)} \mW[C_\xi^{(\cdot)}] - (C^{(k)}_\xi)^2 \right) + \sqrt{\gamma} C_\xi^{(k)}\; \eta(t) \right\}w^{(k)},
\end{eqnarray}
\endnumparts
where $C_\xi^{(k)} = \int dx\; c_\xi(x) \left(|\alpha^{(k)}(x)|^2 - \delta(0)/2\right)$. Observables were calculated from equations~(31) using the weighted average defined in equation~(\ref{eqn:weigtedavg}) and the correspondences (\ref{eqn:expectationValues}).

\begin{figure*}[tb]
\begin{minipage}[b]{0.34\textwidth}
\includegraphics[width=\textwidth]{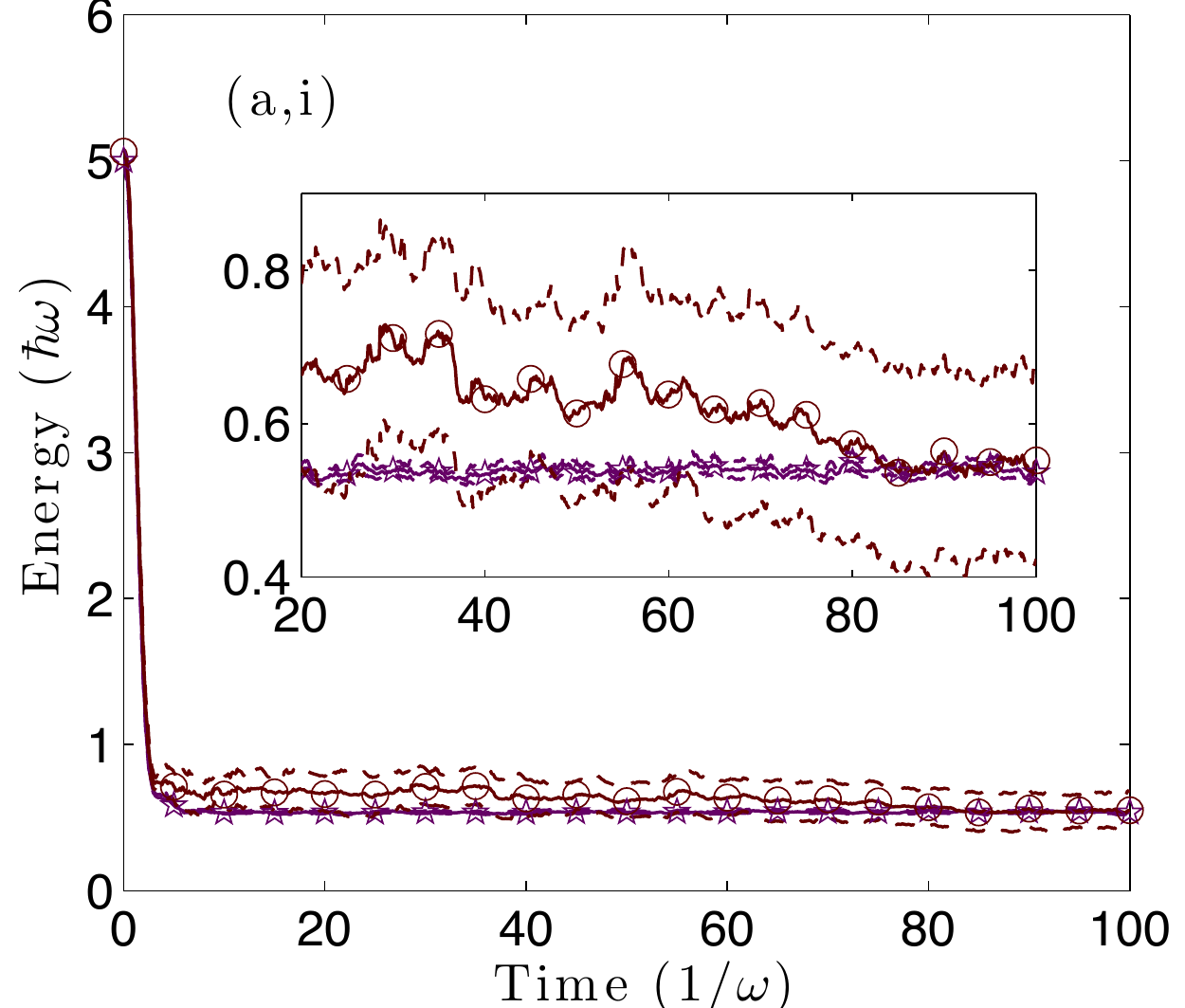}
\includegraphics[width=\textwidth]{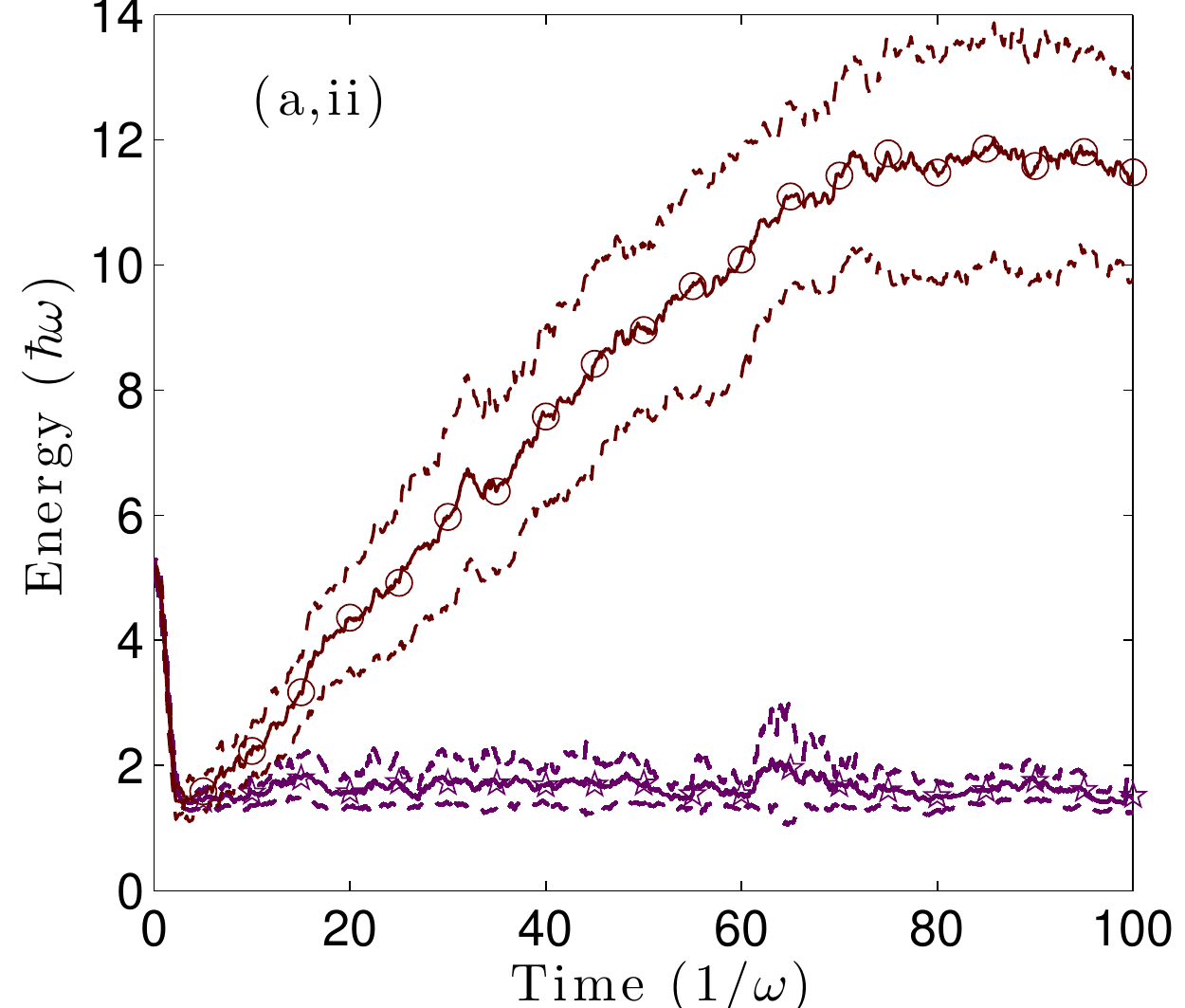}
\end{minipage}
\includegraphics[width=0.01\textwidth]{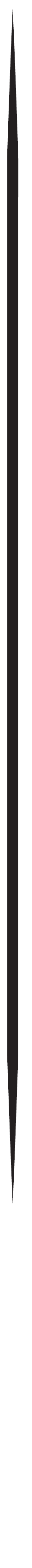}%
\includegraphics[width=0.3\textwidth]{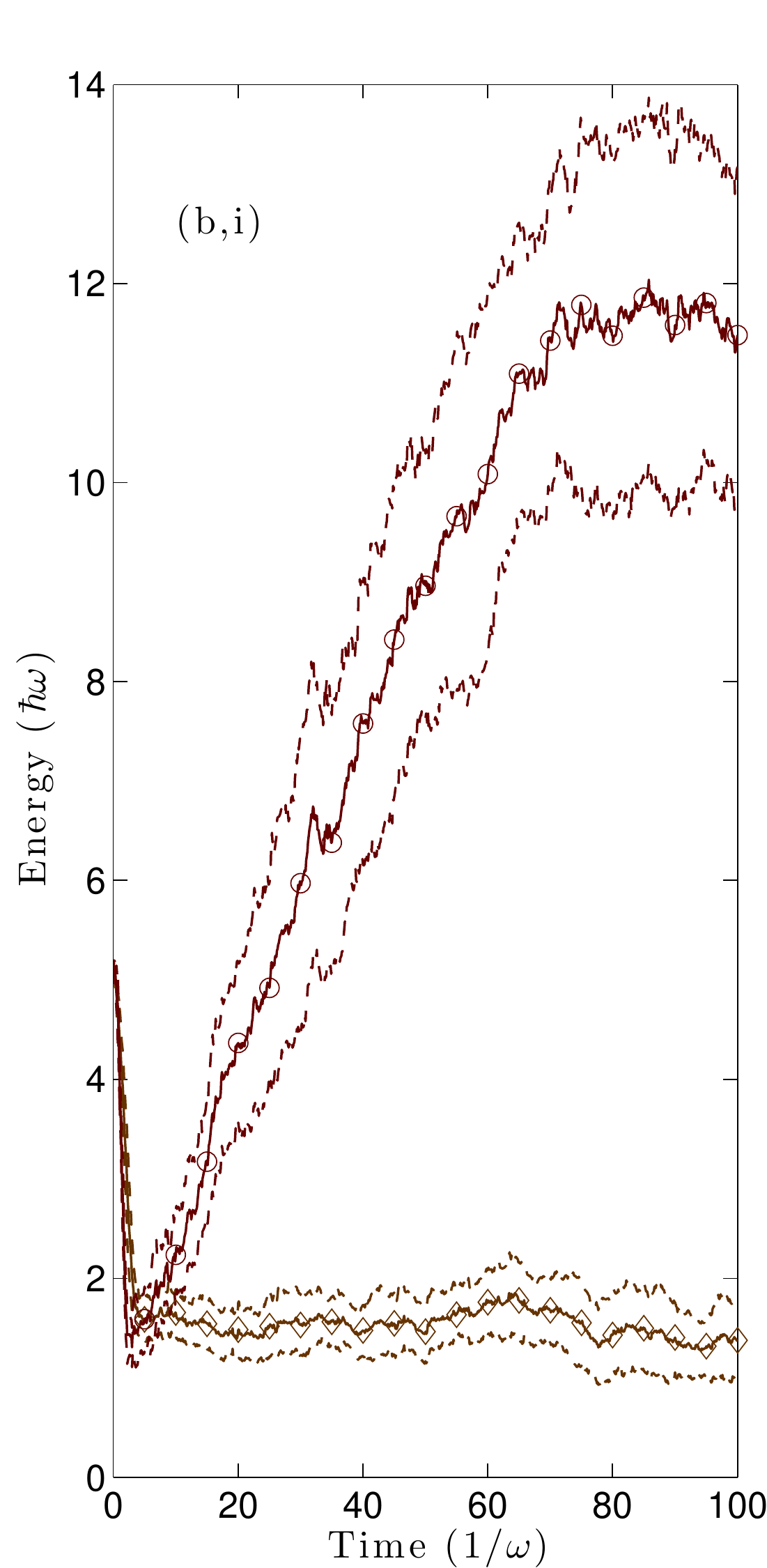}%
\begin{minipage}[b]{0.34\textwidth}
\includegraphics[width=\textwidth]{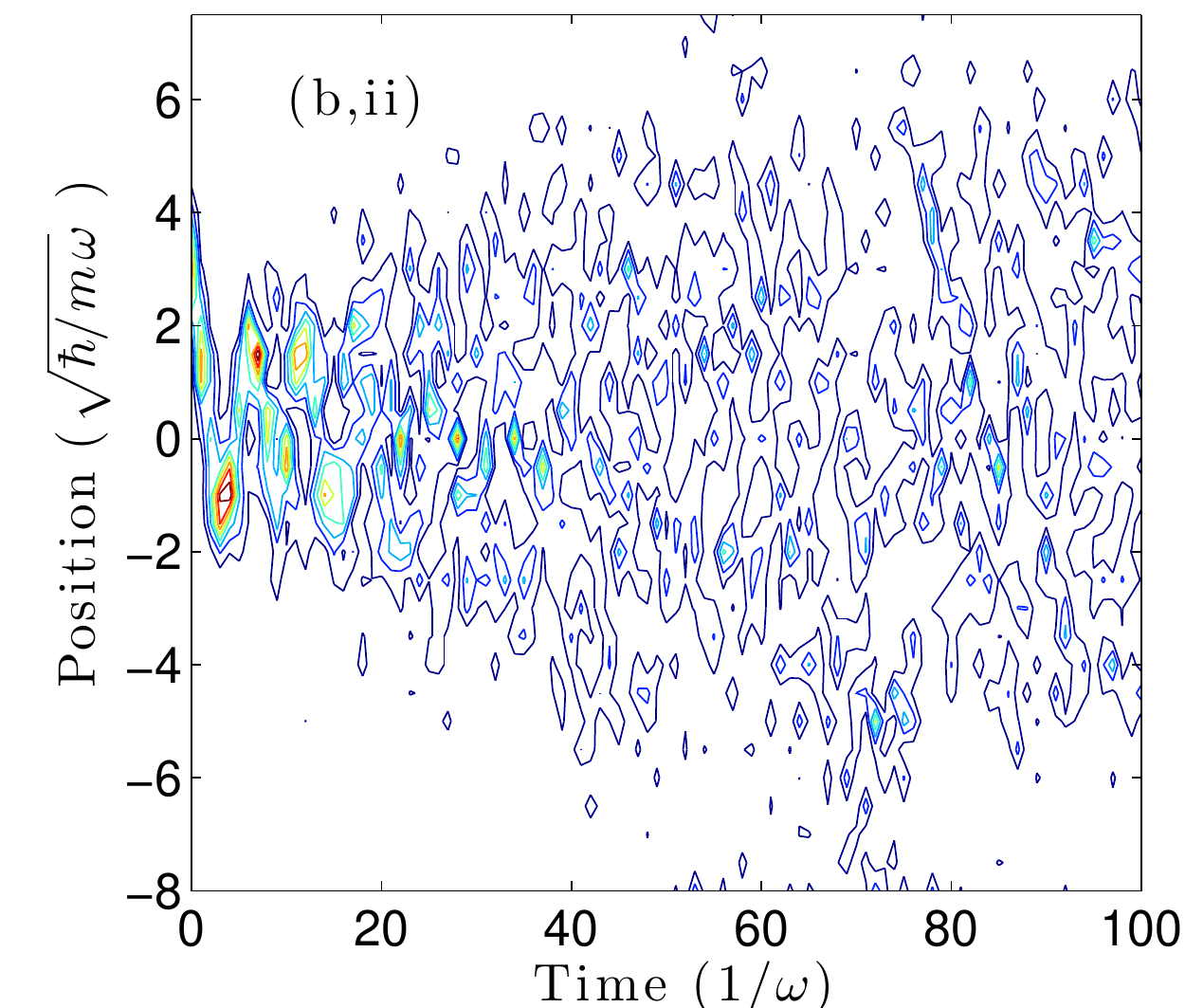} \\ %
\includegraphics[width=\textwidth]{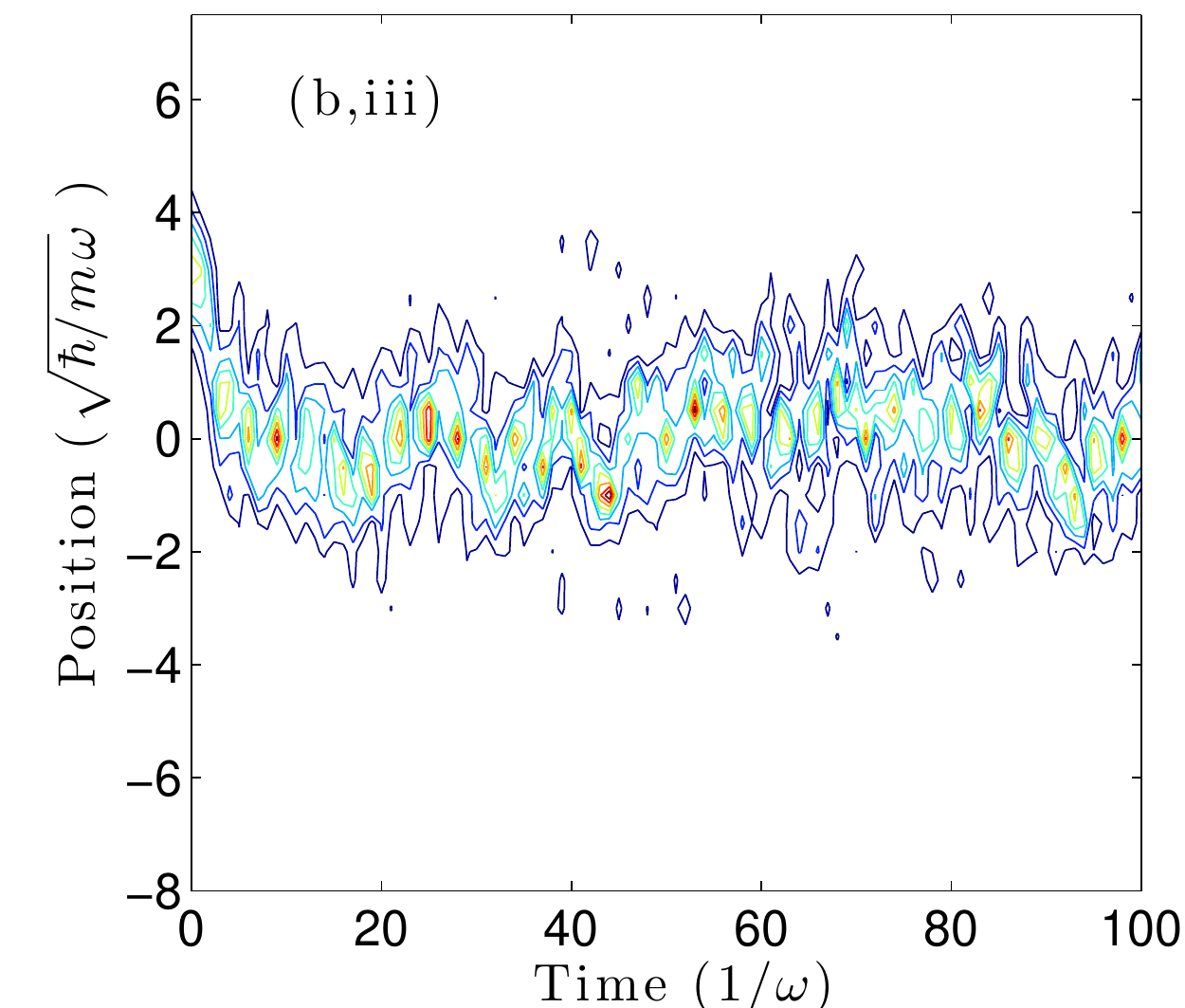}%
\end{minipage}
\caption{(a) Comparison of the Hartree-Fock method and the NPW particle filter for the simulation of a feedback-controlled BEC undergoing a continuous cavity-mediated measurement. The NPW particle filter was integrated using equations~(31), averaged over $P=10$ paths, and is plotted in red with circles. Hartree-Fock simulations were performed by integrating equation~(\ref{eqn:cavBEChart}), averaging over $P=100$ paths, and are plotted in purple with stars. A small value of $\xi = 0.1$ is plotted in (a,i), while a higher strength of $\xi = 0.5$ is plotted in (a,ii). Both plots have $\gamma = 5$, $N=100$ and $u_s = 1$. The Hartree-Fock method only agrees with the NPW particle filter when $\xi = 0.1$ [see inset of (a,i)]; it fails to predict the additional disruption to the BEC, caused by spontaneous-emission noise, when $\xi = 0.5$.
(b) A comparison of a controlled BEC under cavity-mediated measurement without and with the quantum-noise control, averaged over $P=10$ paths. (b,i) shows a comparison of the per-particle energy of the system, while the density fluctuations in the system without and with control are shown in (b,ii) and (b,iii), respectively. In (b,i): integration of the NPW particle filter (31) without quantum-noise control, $u_c = 0$, is plotted in red with circles; integration with the addition of the quantum-noise control, $u_c = 5$, is plotted in orange with diamonds. (b,ii) and (b,iii) are contour plots of the BEC density over an individual path, with (light) blue hues to (dark) red hues indicating low and high density, respectively. The noise in the contour plots is a physical consequence of the measurement process, rather than numerical uncertainty (which is less than 0.1\%). All integrations were performed with $u_s = 1$, $\xi = 0.5$, $\gamma = 5$ and $N=100$.}
\label{fig:cavBECxicomp}
\end{figure*}

A comparison of the Hartree-Fock method and NPW particle filter simulations\footnote{Simulations were completed with the numerical integration package XMDS2 \cite{Dennis:2013}} of a feedback-controlled BEC undergoing a continuous cavity-mediated measurement are shown in figure~\ref{fig:cavBECxicomp}(a). For both the Hartree-Fock wavefunction and NPW particle filter, the initial state was a BEC with a constant phase and amplitude, given by the displaced Gaussian
\begin{equation}
	\alpha_0(x)= \sqrt{\frac{N}{\sqrt{2 \pi} \sigma}}\exp\left(-\frac{(x-x_0)^2}{4\sigma^2}\right), \label{eqn:ICdispgauss}
\end{equation}
where $x_0$ is the Gaussian's displacement from the origin, $\sigma$ is its variance and $N$ is the average total number of the condensate. More details on the choice of initial condition, a technique for sampling this initial state, and a discussion on how it affects the convergence of the numerics can be found in \ref{apx:sampNPWparfil}. When the cosine measurement is close to a position measurement, $\xi=0.1$, we see the NPW particle filter and Hartree-Fock simulations match well, with both predicting damping of the BEC to a steady state. For larger values of $\xi$, the cosine measurement is no longer `position-like', in that it has some curved structure on the length scale of the BEC.  When $\xi=0.5$, we see that the NPW particle filter predicts significantly more disruption to the BEC than the Hartree-Fock solution. This disruption, due to the measurement backaction, causes additional heating that \emph{cannot} be counteracted by simple linear feedback damping. Consequently, if a feedback-control experiment was designed solely on the basis of the Hartree-Fock simulation, then this experiment would likely fail, for it would not account for this additional backaction effect neglected by the Hartree-Fock simulation. 

The Hartree-Fock method misses the additional measurement backaction effect on the BEC because it neglects certain spontaneous-emission events. Recall that the derivation of conditional master equation (\ref{eqn:cavBECmasts}) assumes that the light interacting with the two-level atoms is sufficiently far-detuned from the atomic resonance that the atomic excited state can be adiabatically eliminated. Thus, all spontaneous-emission events in the system are now encoded as scattering events between the light and BEC. These spontaneous-emission events change the phase of the outgoing probe beam, which is how the cosine-squared moment of the BEC is measured. As the Hartree-Fock approximation assumes all atoms are in the same single-particle state, it restricts which spontaneous-emission events can be described.  In contrast, the NPW particle filter allows non-collective states of the atoms, and can therefore simulate \emph{all} spontaneous-emission events. It is for this reason that we call this additional backaction effect \emph{spontaneous-emission noise}. Note also that spontaneous-emission noise arises due to non-classical correlations in the atomic field. Hence, in some sense, it is also a form of quantum noise.

Although the effects of spontaneous-emission noise cannot be removed via simple linear damping, they \emph{can} be cancelled with a more exotic choice of feedback control. We present this control, and demonstrate its effectiveness, in the next section.

\subsection{Quantum-noise control} \label{sec_q_noise_control}
The Hartree-Fock simulation converged to the NPW particle filter result when $\xi=0.1$, which shows that the net effect of the measurement backaction is partially mode-dependent. Whether the BEC cools to a steady state depends on the competition between the heating due to the scattering and the damping due to the feedback.  The linear feedback used in the simulations of figure~\ref{fig:cavBECxicomp}(a) damped the centre-of-mass motion of the BEC, which for $\xi=0.1$ is very close to the spatial mode of the disruption due to measurement backaction. When $\xi=0.5$, energy was added to modes \emph{other} than the position moment mode, so we hypothesized that the solution was to \emph{add an extra control} that targets the \emph{cosine-squared mode} of the BEC. 

A methodology used to damp specific modes of a BEC is described in \cite{Haine:2004}. We can apply this methodology to create a control that targets the cosine-squared, $c_\xi(x)$, mode of the BEC. The single-particle Hamiltonian that includes this new \emph{quantum-noise control} is 
\begin{eqnarray}
h_{FC}(x,u_s,u_c)  =  h_F(x,u_s) + \frac{u_c c_\xi(x)}{\langle \hat{N} \rangle} \int dy \;  c_\xi'(y) \mbox{Im}\left[\langle \hat{\psi}^\dag(y) \hat{\psi}'(y) \rangle\right] , \label{eqn:hamilcavQon}
\end{eqnarray} %
where $u_c$ is the strength of the quantum-noise control for the cavity-mediated measurement, $\hat{\psi}'(x) \equiv \partial_x \hat{\psi}(x)$,  and $c_\xi'(x) \equiv \partial_x c_\xi(x)$.

NPW particle filter simulations of a BEC undergoing continuous cavity-mediated measurement without and with this novel quantum-noise control are shown in figure~\ref{fig:cavBECxicomp}(b). In part (b,i) we can see that the disruption due to the spontaneous-emission noise is completely cancelled by the quantum-noise control.  Comparing parts (b,ii) and (b,iii), we see that the quantum-noise control vastly improves the stability of the BEC under continuous monitoring. In (b,ii) we see that a continuous cavity-mediated measurement causes the BEC to break apart, even in the presence of feedback. In contrast, (b,iii) shows that the additional quantum-noise control cancels the spontaneous-emission noise, thereby preventing this spread. 

We have demonstrated that spontaneous-emission noise can be cancelled in a weakly-interacting BEC undergoing continuous cavity-mediated measurement, thus showing that a BEC can be feedback-cooled to a steady state close to the ground-state energy. However, the conditional master equation (\ref{eqn:cavBECmasts}) that describes the cavity-mediated measurement is only valid when inter-atomic interactions in the condensate are negligible.  Typically, inter-atomic interactions make the BEC larger than the optical wavelength, bringing it out of the near-position measurement regime. Indeed, the coupling between the cavity mode and collective position mode requires negligible inter-atomic interactions. Although a non-interacting condensate can be created with a dilute atomic sample or via a Feshbach resonance \cite{Leggett:2001}, it is technically demanding.  Furthermore, there are instances where large inter-atomic interactions are desirable, such as for stable atom laser operation \cite{Haine:2002,Haine:2003} and for the generation of non-classical atom laser states \cite{Johnsson:2007}. Fortunately, feedback control is still possible in this regime using a phase-contrast imaging setup.

\begin{figure}[tb]
\includegraphics[scale=0.45 ]{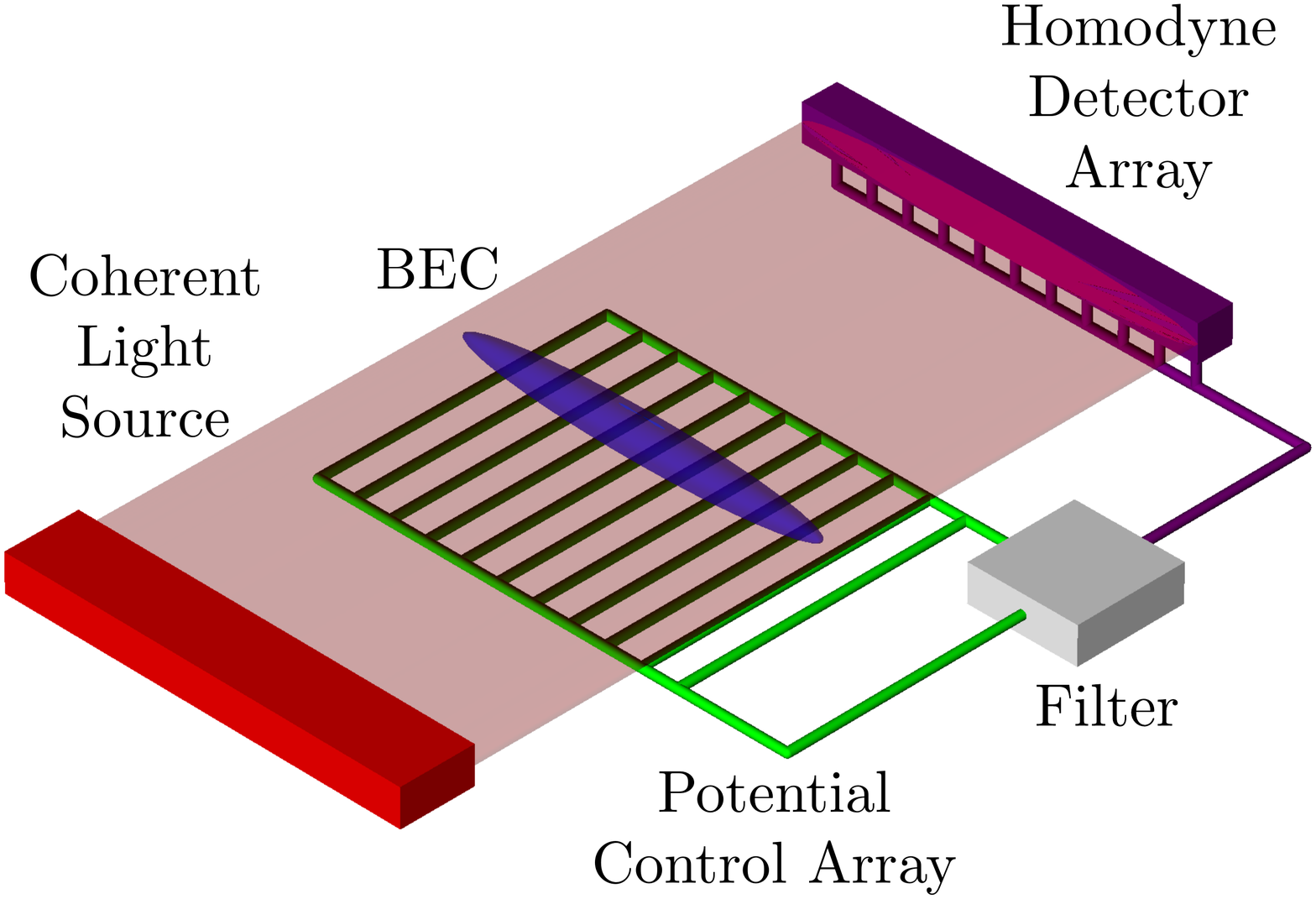}
\caption{Schematic diagrams of a BEC under feedback control using continuous phase-contrast imaging. The condensate is illuminated by off-resonant light. After interacting with the atoms, the light is detected by an array of CCD cameras (which we can model as an array of homodyne measurements). The green wires represent controls that modify the trapping field with the aim of driving the BEC to a low energy steady state, the trapping field itself is not shown.}
\label{fig:directBECdiag}
\end{figure}
\subsection{Phase-contrast imaging}
We consider the feedback control of a BEC with inter-atomic interactions of arbitrary strength $U$ under phase-contrast imaging, which has been realized experimentally and has allowed for the real-time continuous measurement of a BEC \cite{Andrews:1996,Bradley:1997}.  In phase-contrast imaging, an off-resonant laser beam passes through the BEC, gaining a phase shift that is proportional to the atomic column density (see figure~\ref{fig:directBECdiag}). This light is measured by an array of CCD cameras, which we model as an array of homodyne detectors. The conditional master equation for a quasi-one-dimensional `cigar-shaped' BEC under phase-contrast imaging is derived in \cite{Szigeti:2009}, and in Stratonovich form is:
\begin{eqnarray}
	\partial_t \hat{\rho} 	&= -i [\hat{H}_{FN}(u_s),\hat{\rho}] + \gamma \int dx \,\mathcal{D}[\hat{M}_{\nu}(x)] \hat{\rho} \nn \\
					&+ \gamma \int dx \, \mathcal{C}[\hat{M}_{\nu}(x)] \hat{\rho} + \sqrt{\gamma} \int dx \, \mathcal{H}[\hat{M}_{\nu}(x)] \hat{\rho} \; \eta(x,t), \label{eqn:dirBECmasts}
\end{eqnarray}
where 
\begin{equation}
	\hat{H}_{FN}=\hat{H}_{F}+\frac{U}{2}\int dx \;\hat{\psi}^\dagger(x)\hat{\psi}^\dagger(x)\hat{\psi}(x)\hat{\psi}(x)
\end{equation}
is the Hamiltonian for an interacting BEC under linear feedback [see equation~(\ref{field_Ham})], and the measurement operators are
\begin{equation}
	\hat{M}_\nu(x)=(\mu_\nu * \hat{\psi}^\dagger \hat{\psi})(x).
\end{equation}
These operators describe a measurement of the local particle density operator, $\hat{\psi}^\dagger(x) \hat{\psi}(x)$, convolved with a kernel function $\mu_\nu(x)$, at every position $x$. Here we use $*$ to indicate a convolution, which is defined as 
\begin{equation}
	(f*g)(x) \equiv \int_{-\infty}^{\infty}dy f(y) g(x-y).
\end{equation}
The kernel function $\mu_\nu(x)$ depends on the dimensionless \emph{resolution length scale} $\nu = x_\perp/k_0 x_{\text{HO}}^2$, where $x_\perp$ is the size of the condensate in the tight trapping directions, $x_{\text{HO}} = \sqrt{\hbar/(m \omega)}$ for loose trapping frequency $\omega$, and $k_0 = 2\pi/\lambda$ is the wavenumber of the off-resonant light. Typically, $\nu \ll 1$. In the regime where light is predominantly scattered in the direction of the laser beam, the kernel function in Fourier space is given by $\tilde{\mu}_\nu(k) = \exp\left(-\nu k^4\right)$. Since there are multiple measurement channels, indexed by the position $x$, there are also multiple Stratonovich noises, $\eta(x,t)$. Note that, like equation~(\ref{eqn:cavBECmasts}), equation~(\ref{eqn:dirBECmasts}) has been written in harmonic oscillator units, where energy is in units of $\hbar \omega$, position is in units of $x_{\text{HO}}$, and time is in units of $1/\omega$. 

In this system, the measurement signal provides a resolution-limited density measurement of the BEC. Unlike the cavity-mediated measurement, in no limit does the signal provide a simple position measurement, although multiple position moments can be estimated from the image, whose resolution is defined by the parameter $\nu$.  We investigate two values for the resolution length scale: $\nu=10$ and $\nu=0.1$, which provide coarse-resolution and fine-resolution measurements of the BEC density, respectively.

Under the Hartree-Fock approximation, equation~(\ref{eqn:dirBECmasts}) reduces to \cite{Szigeti:2010}:
\begin{eqnarray}
	\partial_t \tilde{\phi}(x,t)  	&=   \Bigg\{-i \Big(h_F(x,u_s) + U(N-1) \frac{| \tilde{\phi}(x,t)|^2}{n_\phi(t)}\Big) \nn \\
						&+ 2 \frac{\gamma}{n_\phi(t)} (\mu_{\nu} * \mu_{\nu} * |\tilde{\phi}|^2)(x,t) + \sqrt{\gamma} (\mu_{\nu} * \eta)(x,t) \Bigg\} \tilde{\phi}(x,t). \label{eqn:dirBEChart}
\end{eqnarray}
The NPW particle filter converts equation~(\ref{eqn:dirBECmasts}) to \cite{Hush:2009,Hush:2012,Hush:2012a}: 
\numparts
\label{eqn:dirBECnpwsdes}
\begin{eqnarray}
\fl	\partial_t \alpha^{(k)}(x) = -i \Big( h_F(x,u_s) + U |\alpha^{(k)}(x)|^2 + \sqrt{\gamma} (\mu_{\nu} * \zeta^f)(x) \Big) \alpha^{(k)}(x), \\
\fl	\partial_t w^{(k)} =  2 \int dx  \Big( \gamma \big( 2 M_\nu^{(k)}(x) \mW[M_\nu^{(\cdot)}(x)]  - M_\nu^{(k)}(x)^2 \big)  + \sqrt{\gamma} M_\nu^{(k)}(x,t)\; \eta(x) \Big)w^{(k)}, 
\end{eqnarray}
\endnumparts
where $M_\nu^{(k)}(x) = (\mu_\nu * |\alpha^{(k)}|^2)(x)$.   

\begin{figure*}[h!tb]
\begin{minipage}[b]{0.34\textwidth}
\includegraphics[width=\textwidth]{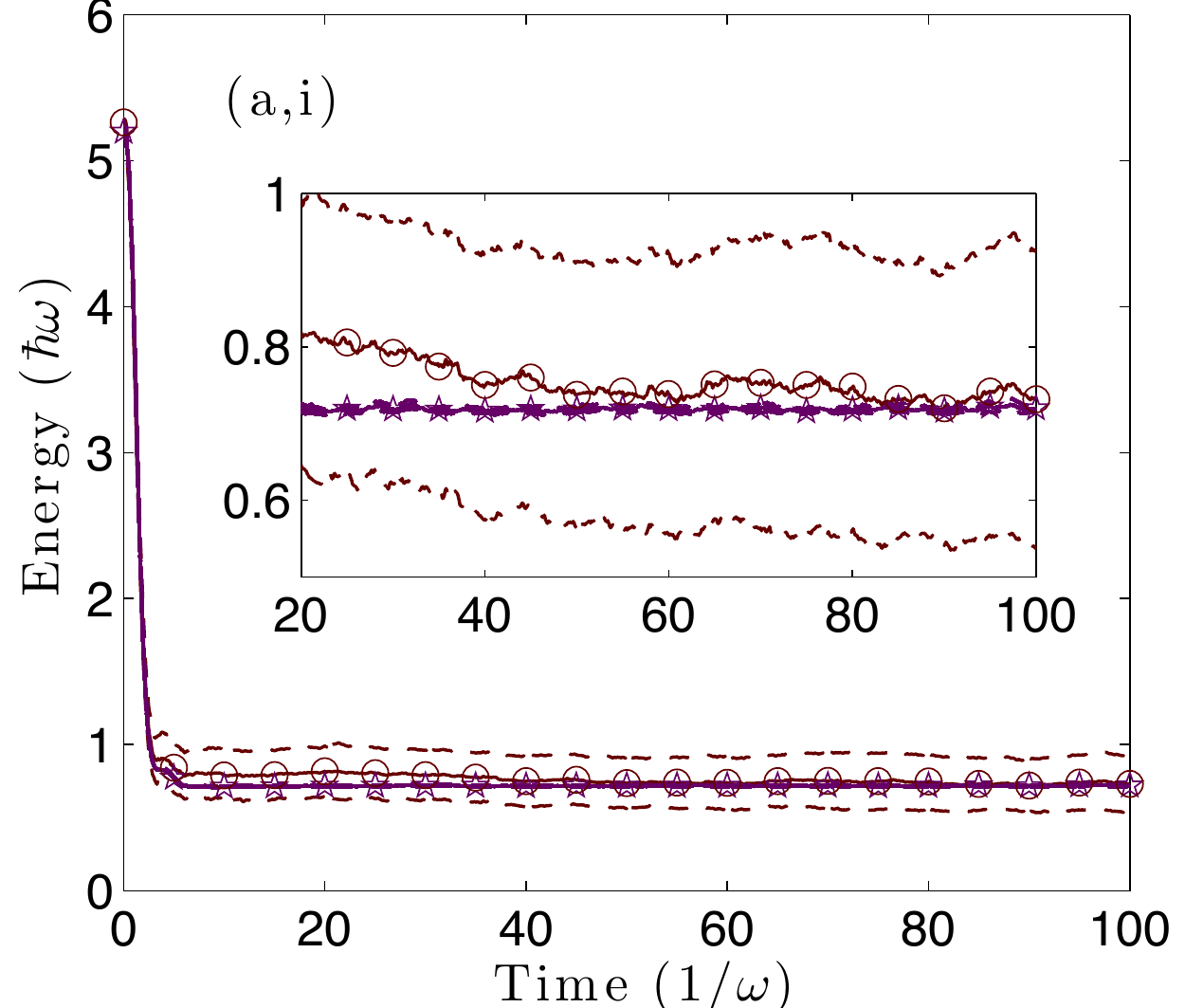}
\includegraphics[width=\textwidth]{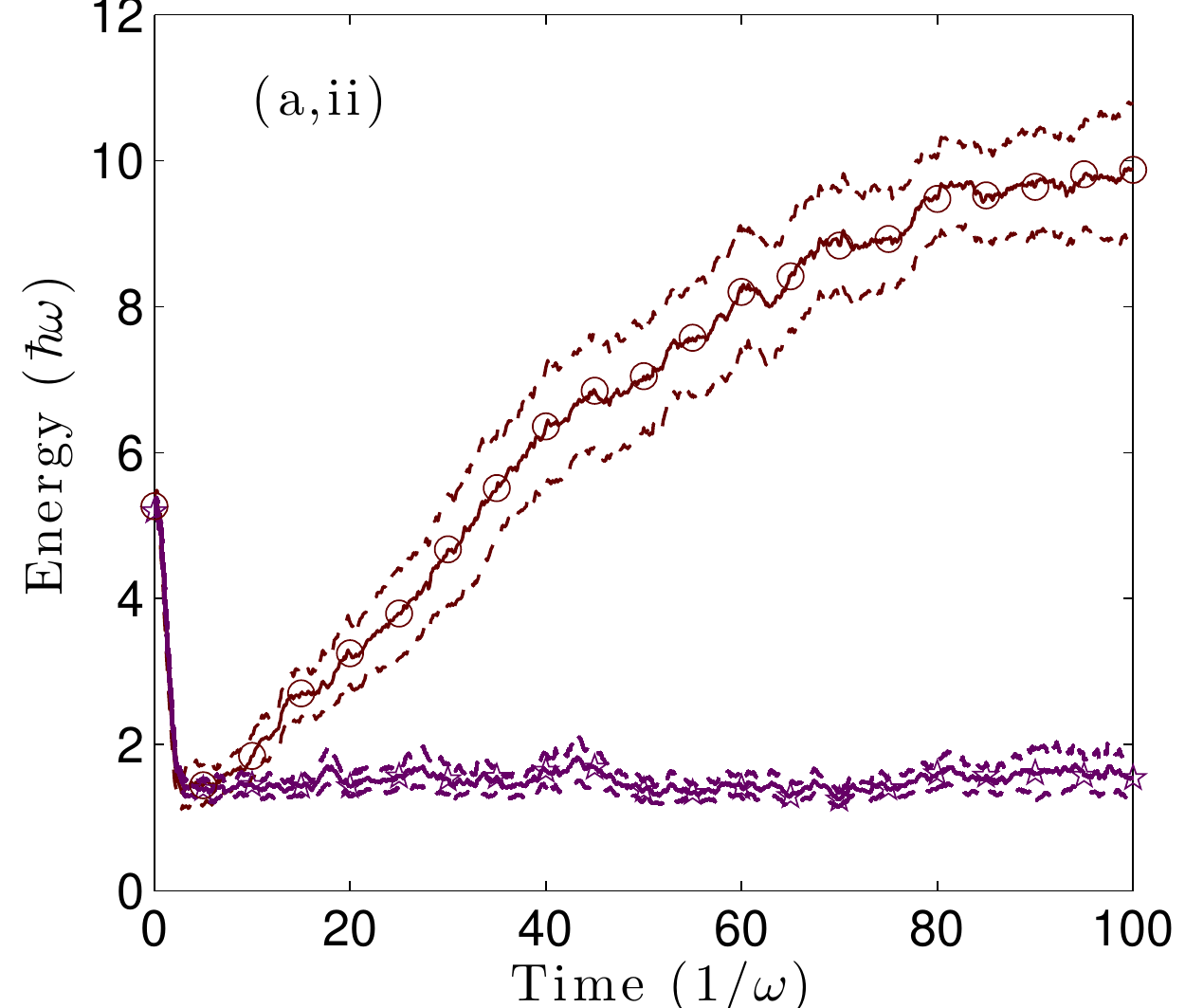}
\end{minipage}
\includegraphics[width=0.01\textwidth]{VerticleLine.pdf}%
\includegraphics[width=0.3\textwidth]{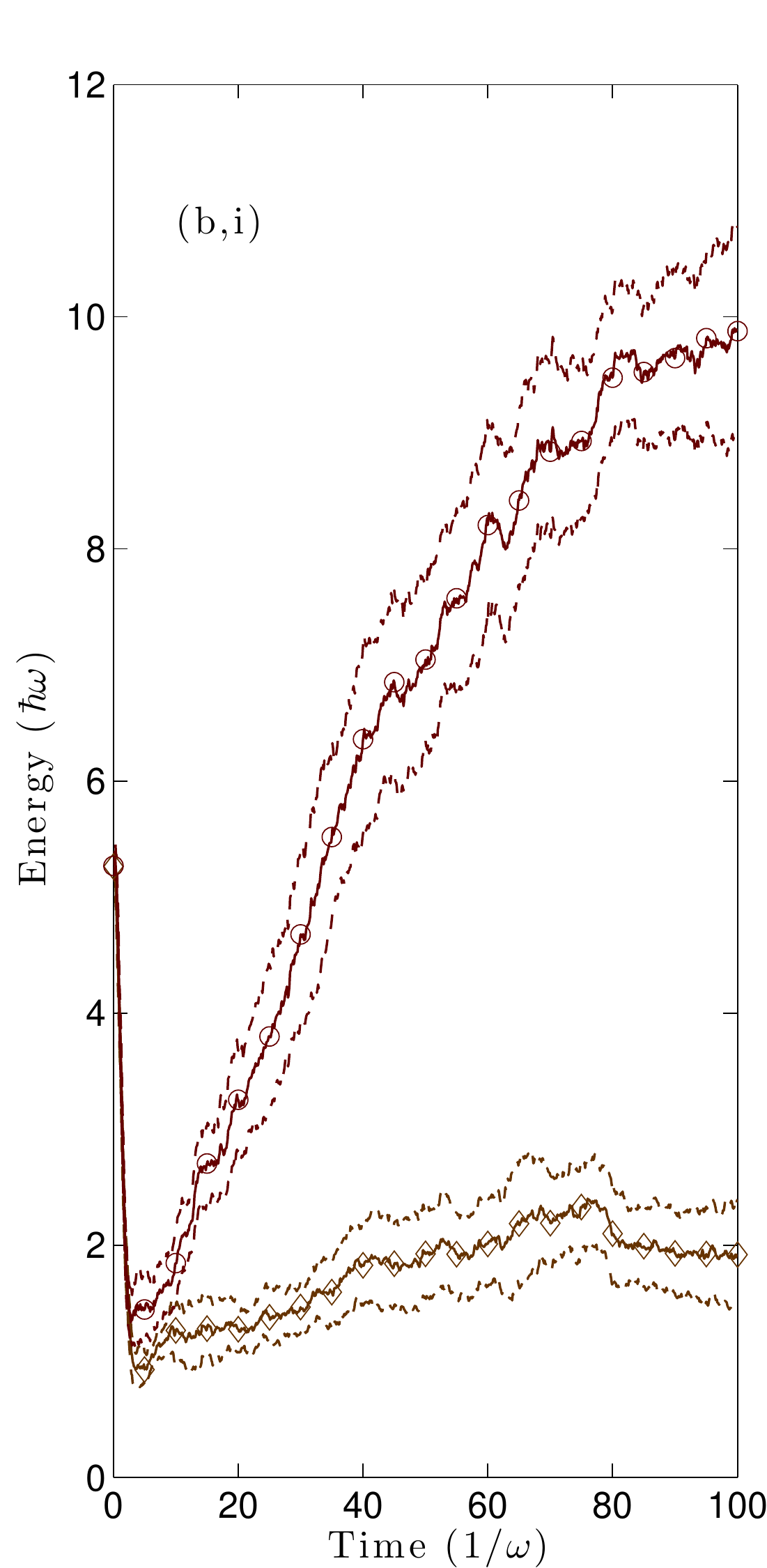}%
\begin{minipage}[b]{0.34\textwidth}
\includegraphics[width=\textwidth]{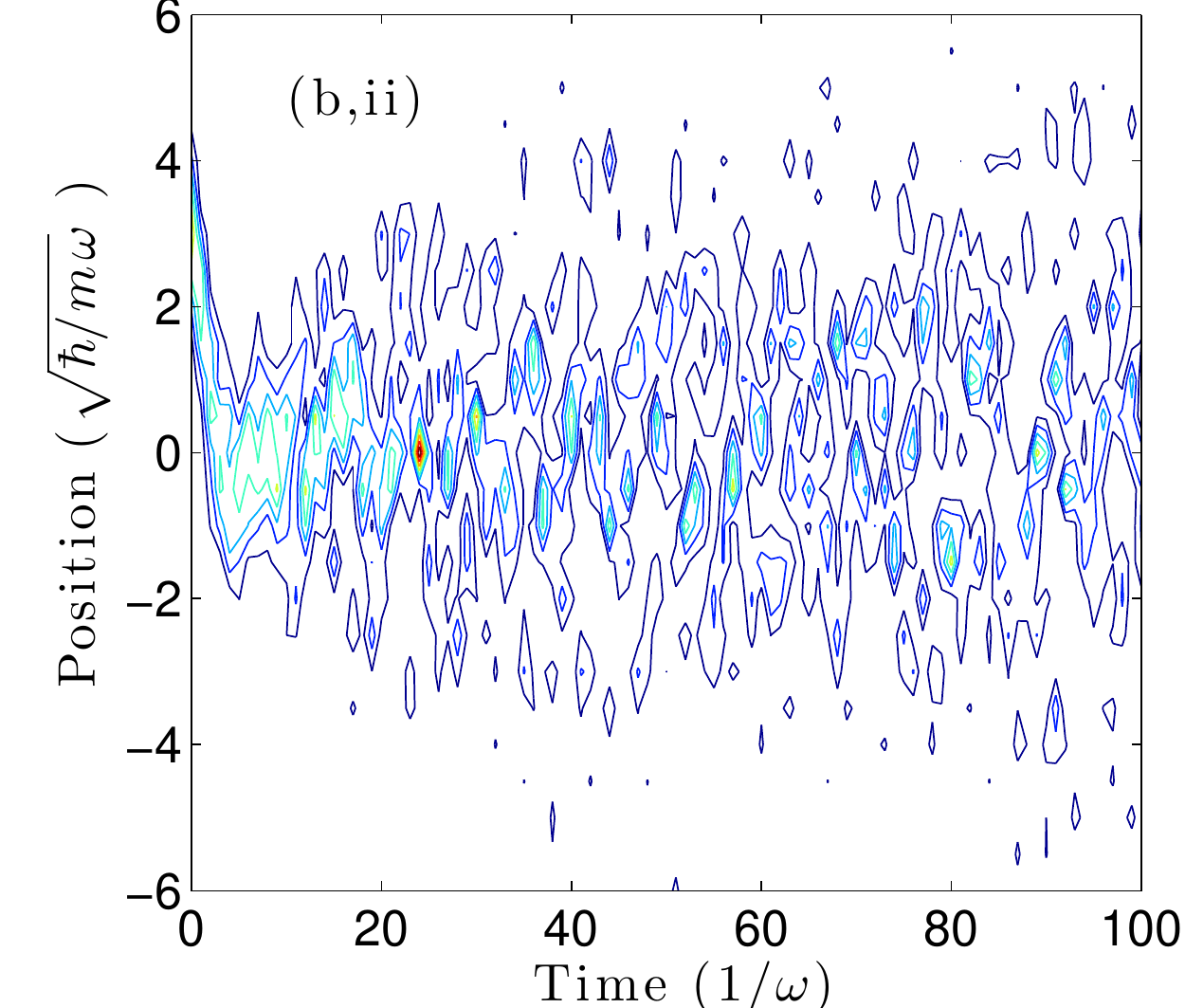} \\ %
\includegraphics[width=\textwidth]{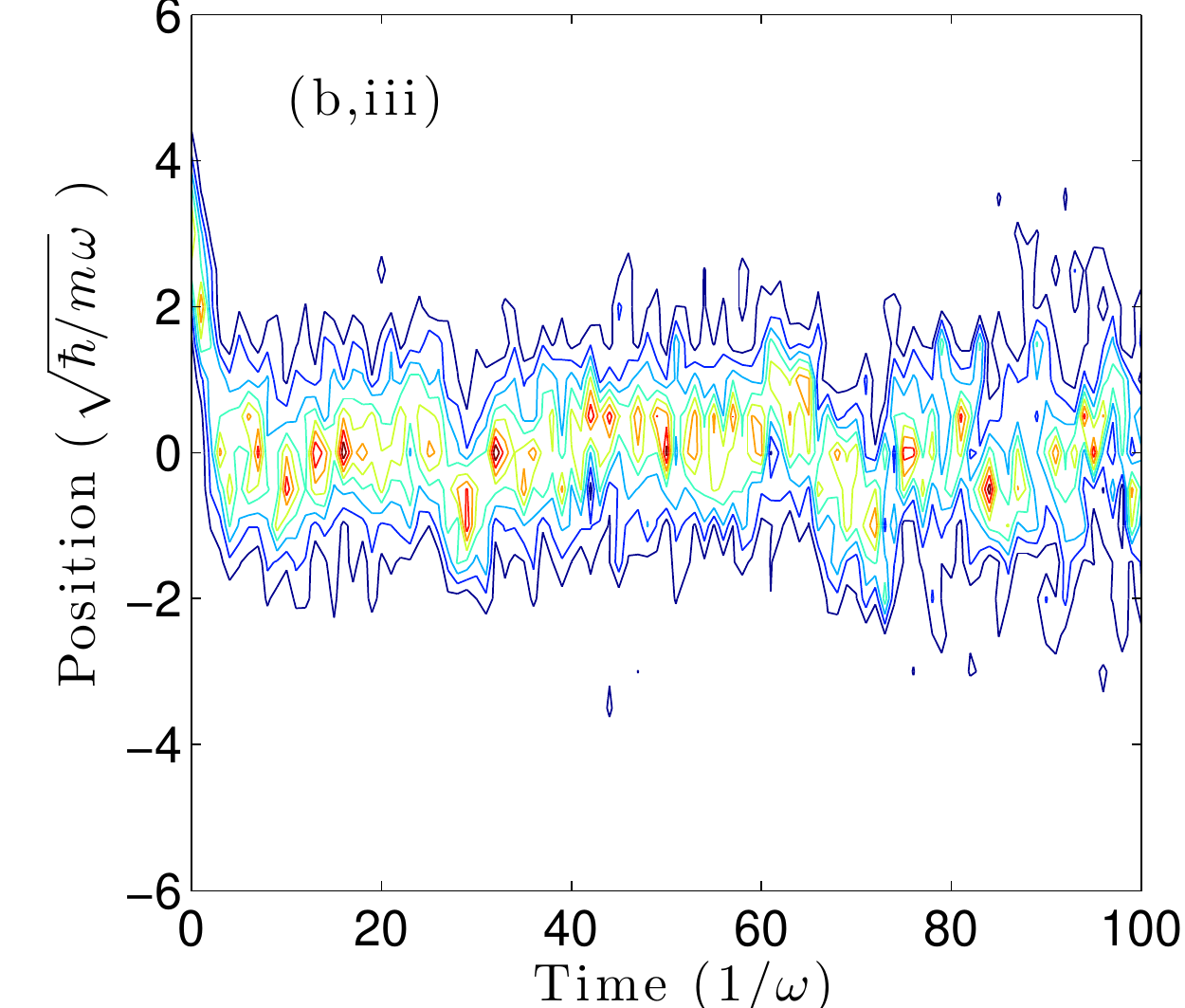}%
\end{minipage}
\caption{(a) Comparison of the Hartree-Fock method and the NPW particle filter for the simulation of a feedback-controlled BEC undergoing continuous phase-contrast imaging. The NPW particle filter was integrated using equations~(39), averaged over $P=10$ paths, and is plotted in red with circles. Hartree-Fock simulations were performed by integrating equation~(\ref{eqn:dirBEChart}), averaging over $P=100$ paths, and are plotted in purple with stars. A coarse-resolution measurement, $\nu = 10$, is plotted in (a,i), while a fine-resolution measurement, $\nu = 0.1$, is plotted in (a,ii). Both plots have $\gamma = 1$, $N=100$, $U/N = 3$ and $u_s = 1$. The Hartree-Fock method only agrees with the NPW particle filter when $\nu = 10$ [see inset of (a,i)]; it fails to predict the additional disruption to the BEC, caused by spontaneous-emission noise, when $\nu = 0.1$.
(b) A comparison of a controlled BEC under continuous phase-contrast imaging without and with the quantum-noise control averaged over $P=10$ paths. (b,i) shows a comparison of the per-particle energy of the system, while the density fluctuations in the system without and with control are shown in (b,ii) and (b,iii), respectively. In (b,i), the integration of the NPW particle filter (39) without quantum-noise control, $u_\mu = 0$, is plotted with red circles. A simulation of equations~(39) with the addition of the quantum-noise control, $u_\mu = 5$, is plotted in orange with diamonds. (b,ii) and (b,iii) are contour plots of the density of the BEC over an individual path, with (light) blue hues to (dark) red hues indicating low to high densities, respectively. The noise in the contour plots is a physical consequence of the measurement process, rather than numerical uncertainty (which is less than 0.1\%). All integrations were performed with $u_s = 1$, $\nu = 0.1$, $U/N = 3$, $\gamma = 1$ and $N=100$.}
\label{fig:dirBECnucomp}
\end{figure*}

A comparison of Hartree-Fock method and NPW particle filter simulations of equation~(\ref{eqn:dirBECmasts}) is shown in figure~\ref{fig:dirBECnucomp}(a). This analysis is performed \emph{without} a quantum-noise control.  When $\nu = 10$ we see that the Hartree-Fock method agrees with the NPW particle filter, and both predict cooling of the BEC to a steady state. This shows that for coarse-resolution density measurements, higher-order modes are weakly excited such that feedback to the position mode provides sufficient damping, and the energy of the BEC stays low.  Unfortunately, reaching this parameter limit can be experimentally difficult \cite{Szigeti:2010}.  When the BEC is probed at a finer spatial resolution ($\nu = 0.1$), the Hartree-Fock solution \emph{diverges} from the NPW particle filter solution, which shows that the spontaneous-emission noise excites higher-order modes of the BEC that are not cooled by the linear feedback control. 

As with the cavity-mediated measurement, we can introduce controls for higher-order modes that cancel the heating caused by spontaneous-emission noise.  Since our model of phase-contrast imaging gives a multi-channel measurement, we require multiple channels of feedback, effected via the trapping potential. The single-particle Hamiltonian $h_{FM}(x,u_s,u_\mu)$ describing this new feedback is: 
\begin{eqnarray}
	h_{FM} (x,u_s,u_\mu) = h_F(x,u_s) + \frac{u_\mu}{\langle \hat{N} \rangle}\left(\mu_\nu *  \mu_\nu' * \mbox{Im}[\langle \hat{\psi}^\dag \hat{\psi}' \rangle]\right)(x) , \label{eqn:hamildirQon}
\end{eqnarray}
where $\mu_\nu'(x) = \partial_x \mu_\nu(x)$ and $u_\mu$ is the strength of the quantum-noise control for the direct measurement. Note that this feedback requires a high degree of spatial and temporal control of the BEC's trapping potential. Fortunately, there are several proposals for implementing such potentials experimentally \cite{Fatemi:2007,Rhodes:2006,Pasienski:2008,Nedjalkov:2004a,Bruce:2011}. 

The effectiveness of the quantum-noise control defined in equation~(\ref{eqn:hamildirQon}) is shown in figure~\ref{fig:dirBECnucomp}(b). In (b,i) we can see that the quantum-noise control completely cancels the disruption caused by a strong phase-contrast measurement of the BEC.  In parts (b,ii) and (b,iii) we compare the density of the BEC without and with the quantum-noise control, respectively.  In (b,ii) we see that the spontaneous-emission noise causes the BEC to spread rapidly.  In contrast, (b,iii) shows that the quantum-noise control (\ref{eqn:hamildirQon}) prevents the density excitations that cause the condensate to break apart. 

\subsection{Discussion of Experimental Feasibility}

We have shown that feedback control can be used to stabilize a continuously-monitored BEC close to the ground-state energy for both measurement apparatuses shown in figure~\ref{fig:cavBECdiag} and figure~\ref{fig:directBECdiag}, and for a wide range of parameter regimes. Experimentally, it is simplest to implement \emph{only} linear feedback control via adjustments to the trapping minimum\footnote{Quadratic control, via adjustments to the trapping frequency, are also relatively simple to implement. However, like linear control, quadratic control cannot cancel spontaneous-emission noise, and so could only be used in the weak-probing regime.}. It might therefore be tempting to try and feedback-cool a BEC in the `weak-probing' regime (e.g. $\xi = 0.1$ for the cavity-mediated measurement and $\nu = 10$ for phase-contrast imaging), since spontaneous-emission noise is negligible and only the bulk feedback mechanism, governed by parameter $u_s$, is required to achieve net cooling to steady state. However, operating in this parameter regime requires relatively small, tightly-trapped condensates \cite{Doherty:1999,Szigeti:2009}, although this may change with the development of trapping technologies. If the target condensate does not satisfy these requirements, more complex measurement and feedback is required. For moderately-sized condensates, a cavity-mediated measurement with an extra quantum-noise control appears promising. As demonstrated in section~\ref{sec_q_noise_control}, the shape of the quantum-noise control must match the measurement. Thus, in the case of a continuous cavity-mediated measurement, it might be possible to implement the quantum-noise control via the probe beam itself. For very large, loosely-trapped condensates, it is likely phase-contrast imaging will be the only option, in which case full spatial control of the trapping potential will be required.

\section{Conclusion}
 
This paper has investigated the previously unexplored multi-mode quantum-field dynamics of a feedback-cooled multi-mode BEC undergoing (a) continuous cavity-mediated cosine-squared measurement and (b) continuous phase-contrast imaging. Extended quantum-field simulations, performed with the recently developed NPW particle filter, have revealed regimes with additional measurement-induced disruption to the condensate that are \emph{not} included in single-mode and semiclassical models, and furthermore cannot be counteracted with simple linear and/or quadratic feedback controls. Fortunately, we have developed a more sophisticated quantum-noise control that cancels the effect of this additional measurement backaction, and shown that this allows for successful feedback-cooling of the condensate to a steady state. This has been an important demonstration of the necessity of multi-mode quantum-field simulations, and in particular the NPW particle filter, to the successful design and implementation of measurement-based feedback-control schemes for multi-mode quantum systems.

\section*{Acknowledgements}

Simulations were completed using the NCI National Facility in Canberra, Australia, which is supported by the Australian Commonwealth Government. J. J. Hope would like to acknowledge the support of the ARC Future Fellowship programme. A. R. R. Carvalho gratefully acknowledges support by the Australian Research Council Centre of Excellence for Quantum Computation and Communication Technology (Project No. CE110001027) and S. S. Szigeti acknowledges support by the Australian Research Council Centre of Excellence for Engineered Quantum Systems (Project No. CE110001013)

\appendix

\section{Derivation of the NPW particle filter} \label{apx:derivNPWparfil}

The derivation of the NPW particle filter (15), which is used to simulate the conditional master equation~(\ref{eqn:conBECmasts}), is performed in three steps. Firstly, the evolution of the conditional master equation is transformed exactly to that of a \emph{functional quasi-probability} distribution, specifically the NPW representation \cite{Hush:2010}. Secondly, we use a series of approximations validated in \cite{Hush:2012} to express the evolution of the NPW quasi-probability distribution in the form of a Kushner-Stratonovich equation. Finally, we translate the quasi-probability distribution evolution to that of a set of weighted stochastic differential equations (WSDEs) using \cite{Hush:2009}. The similarity of the integration methods between the WSDEs and particle filters is why we termed this final result a NPW particle filter in the main text. The advantage of the NPW particle filter is that it allows for an approximate simulation of the conditional master equation (\ref{eqn:conBECmasts}) that includes interesting and important quantum correlations in the atomic field, yet still scales more favourably than direct integration of the conditional master equation (see comment on curse of dimensionality in section~\ref{sec:condmasteqn}).

It is convenient to express equation~(\ref{eqn:conBECmasts}) in terms of the number operator and phase operator (as defined by Susskind and Glowgower \cite{Susskind:1964}) before transforming to the NPW representation. For most of the terms in equation~(\ref{eqn:conBECmasts}) this is straightforward, as we simply use the definition of the local number operator $\hat{n}(\mv{x}) = \hat{\psi}^\dag(\mv{x}) \hat{\psi}(\mv{x})$. The most difficult term is the single-particle Hamiltonian $h(\mv{x},\mv{u})$, as it may contain derivatives and so we cannot simply assume that it commutes with the field operators. Instead, we define the functional $h'(\mv{x},\mv{y},\mv{u}) = h(\mv{y},\mv{u})\delta(\mv{x} - \mv{y})$, giving
\begin{eqnarray}
\int d\mv{x} \, \hat{\psi}^\dag(\mv{x}) h(\mv{x},\mv{u}) \hat{\psi}(\mv{x}) = \int d\mv{x} \,d\mv{y}\, h'(\mv{x},\mv{y},\mv{u})  \hat{\psi}^\dag(\mv{x}) \hat{\psi}(\mv{y}), 
\end{eqnarray}
where $h'(\mv{x},\mv{y},\mv{u})$ \emph{does} commute with $\hat{\psi}(\mv{x})$ and $\hat{\psi}^\dag(\mv{y})$. We now change the field operators to number and phase operators using the identities 
\begin{equation}
	\hat{\psi}(\mv{x}) = \sqrt{\hat{n}(\mv{x}) + \delta(0)}e^{i\hat{\Phi}(\mv{x})} = e^{i\hat{\Phi}(\mv{x})}\sqrt{\hat{n}(\mv{x})}.
\end{equation}
These correspondences are used to transform the Hamiltonian and measurement operators of equation~(\ref{eqn:conBECmasts}) to
\begin{eqnarray}
	\fl \hat{H} =  \int d\mv{x} \, d\mv{y} \; \Big( h'(\mv{x},\mv{y},\mv{u}) \sqrt{\hat{n}(\mv{x})}(e^{i\hat{\Phi}(\mv{x})})^\dag \sqrt{\hat{n}(\mv{y})+\delta(0)}e^{i\hat{\Phi}(\mv{x})} \Big) + \frac{U}{2}\int d\mv{x} \;(\hat{n}(\mv{x}))^2, \label{eqn:modhamillossA}\\
	\fl \hat{L}_i  = \int d\mv{x} \; l_i(\mv{x}) \hat{n}(\mv{x}), \label{eqn:modhamillossB}
\end{eqnarray}
respectively, which is correct up to a constant offset in the single-particle Hamiltonian, $h(\mv{x},\mv{u})$.

The next step is to translate the evolution of the conditional density matrix, $\hat{\rho}$, in equation~(\ref{eqn:conBECmasts}) to the evolution of a functional NPW distribution $\sN[n(\cdot),\varphi(\cdot)]$. We define the functional NPW representation by applying the single-mode NPW representation defined in \cite{Hush:2010} to every point in space:
\begin{eqnarray}
	\fl \sN[n(\cdot),\varphi(\cdot)] = \sum_{k(\cdot)=-n(\cdot)}^{n(\cdot)} \Tr \Big[ \hat{\rho}\prod_\mv{x}\Big( \frac{e^{-2i\phi(\mv{x})k(\mv{x})}}{2\pi}| n(\mv{x}) - k(\mv{x}) \rangle \langle n(\mv{x}) + k(\mv{x})| \Big) \Big],
\end{eqnarray}
where $\sum_{k(\cdot)=-n(\cdot)}^{n(\cdot)}$ denotes a sum over $k(\mv{x})$ from $-n(\mv{x})$ to $n(\mv{x})$ for every point $\mv{x}$ and $\prod_\mv{x}$ denotes a product over every point $\mv{x}$. A functional distribution is required because we are simulating a quantum field. For example, the quasi-probability that there are $n_0$ particles with phase $\varphi_0$ at a spatial point $\mv{x}$ is given by $\sN[n_0 \delta(\cdot), \varphi_0 \delta(\cdot) ]$. For additional details on functional quasi-probability distributions, see \cite{Graham:1970,Daniell:1919,Gardiner:2004}. Finding the evolution of $\sN[n(\cdot),\varphi(\cdot)]$ is achieved using correspondences derived in \cite{Hush:2010}. We present the functional version of these correspondences below: 
\begin{eqnarray}
	\hat{\rho} \hat{n}(\mv{x}) 	\rightarrow \left(n(\mv{x}) - \frac{i}{2} \partial_{\varphi(\mv{x})}\right) \sN[n(\cdot),\varphi(\cdot)], \label{eqn:npwcorres_A}\\ %
	\hat{n}(\mv{x}) \hat{\rho} 	\rightarrow \left(n(\mv{x}) +\frac{i}{2} \partial_{\varphi(\mv{x})}\right) \sN[n(\cdot),\varphi(\cdot)], \\
	\hat{\rho} e^{i{\hat{\Phi}(\mv{x})}} 	\rightarrow e^{i\varphi(\mv{x})} \sN\left[n(\cdot)-\frac{1}{2}\delta(\cdot-x),\varphi(\cdot)\right], \\ %
(e^{i\hat{\Phi}(\mv{x})})^\dag \hat{\rho} 	\rightarrow  e^{-i\varphi(\mv{x})} \sN\left[n(\cdot)-\frac{1}{2}\delta(\cdot-x),\varphi(\cdot)\right], \\ %
	\hat{\rho}  (e^{i\hat{\Phi}(\mv{x})})^\dag 	\rightarrow  e^{-i\varphi(\mv{x})} \Bigg( \sN\left[n(\cdot)+\frac{1}{2}\delta(\cdot-x),\varphi(\cdot)\right]\nn \\ %
	\quad - \int_0^{2\pi} d\varphi'(\cdot) e^{-2i(\varphi(\mv{x}) - \varphi'(\cdot))(n(\mv{x})+\frac{1}{2})} \sN\left[n(\cdot)+\frac{1}{2}\delta(\cdot-x),\varphi'(\cdot)\right] \Bigg), \\ %
	e^{i\hat{\Phi}(\mv{x})} \hat{\rho} 	\rightarrow e^{i\varphi(\mv{x})} \Bigg( \sN\left[n(\cdot)+\frac{1}{2}\delta(\cdot-x),\varphi(\cdot)\right] \nn \\ %
	\quad - \int_0^{2\pi} d\varphi'(\cdot) e^{2i(\varphi(\mv{x}) - \varphi'(\cdot))(n(\mv{x})+\frac{1}{2})} \sN\left[n(\cdot)+\frac{1}{2}\delta(\cdot-x),\varphi'(\cdot)\right] \Bigg),  \label{eqn:npwcorres_F}%
\end{eqnarray}
where $\partial_{\varphi(\mv{x})}$ is a functional partial derivative and $d\varphi(\cdot)$ is a functional integration measure. 

Applying the correspondences (\ref{eqn:npwcorres_A})-(\ref{eqn:npwcorres_F}) to equation~(\ref{eqn:conBECmasts}) using the modified operators (\ref{eqn:modhamillossA}) and (\ref{eqn:modhamillossB}) gives the following evolution for the functional NPW representation:
\begin{eqnarray}
	\fl d \sN[n(\cdot),\varphi(\cdot)] = \Bigg( \frac{-i}{\hbar}  \int d\mv{x} \int d\mv{y} h(\mv{x},\mv{y},\mv{u}) \sqrt{n(\mv{x}) + \frac{i}{2} \partial_{\varphi(\mv{x})} } e^{-i\varphi(\mv{x})} \sqrt{n(\mv{y}) + \delta(0)+ \frac{i}{2}\partial_{\varphi(\mv{y})}} e^{i\varphi(\mv{y})} \nn \\
	\fl \quad \times \Bigg( \sN\left[n(\cdot) - \frac{1}{2} \delta(\cdot - x) + \frac{1}{2} \delta(\cdot - \mv{y}),\varphi(\cdot)  \right] - \int d\varphi'(\cdot) e^{2i(\varphi(\mv{y}) - \varphi'(\cdot))(n(\mv{y}) + 1/2)} \nn \\
	\fl \quad \times \,\sN\left[n(\cdot) - \frac{1}{2} \delta(\cdot - x) + \frac{1}{2} \delta(\cdot - \mv{y}),\varphi'(\cdot)  \right] \Bigg) + \mbox{c.c.} \Bigg) \nn \\
	\fl \quad + \Bigg\{ U \int d\mv{x} \; \partial_{\varphi(\mv{x})} n(\mv{x}) + \sum_i \Bigg( \frac{1}{2} \int d\mv{x} \int d\mv{y} \; l_i(\mv{x})l_i(\mv{y}) \partial_{\varphi(\mv{x})} \partial_{\varphi(\mv{y})} \nn \\
	 \fl \quad - 2 ( (L_i^n)^2 - \sE[(L_i^n)^2] - 2 (L_i^n - \sE[L_i^n])\sE[L_i^n]) + 2  (L_i^n - \sE[L_i^n]) \; \eta_i(t) \Bigg) \Bigg\}\sN[n(\cdot),\varphi(\cdot)],  \label{eqn:conBECnpw}
\end{eqnarray}
where c.c. denotes the complex conjugate,
\begin{equation}
	L_i^n = \int d\mv{x} \; l_i(\mv{x}) n(\mv{x}),
\end{equation}
and
\begin{eqnarray}
	\sE[ f[n(\cdot),\varphi(\cdot)] ] = \int_0^{2\pi} d\varphi(\cdot) \sum_{n(\cdot)=0}^{\infty} f[n(\cdot),\varphi(\cdot)] \sN[n(\cdot),\varphi(\cdot)]
\end{eqnarray}
is our notation for an expectation using the NPW quasi-probability distribution, where $\sum_{n(\cdot)=0}^{\infty}$ denotes a sum over $n(\mv{x})$ for every point $\mv{x}$. The evolution presented in equation~(\ref{eqn:conBECnpw}) is exactly the same as that given by conditional master equation (\ref{eqn:conBECmasts}), as we have not yet applied any approximations. There is also a one-to-one mapping between $\hat{\rho}$ and $\sN[n(\cdot),\varphi(\cdot)]$, but in practice we rarely need to reconstruct $\hat{\rho}$ entirely. Instead, a much more useful connection between the two is through the generation of moments. In \cite{Hush:2010} it was shown that anti-normally ordered field operators are related to NPW expectations by 
\begin{equation}
	\langle (\hat{\psi}(\mv{x}))^q(\hat{\psi}^\dag(\mv{x}))^p \rangle = \sE\left[ \frac{(n(\mv{x})+\frac{p+q}{2} \delta(0))! e^{i(q-p)\varphi(\mv{x})} }{\sqrt{(n(\mv{x})+\frac{q-p}{2}\delta(0))!(n(\mv{x})+\frac{p-q}{2}\delta(0))!}} \right]. \label{eqn:npqmastequiv}
\end{equation}

The solution to the equation of motion for the quasi-probability distribution [equation~(\ref{eqn:conBECnpw})] gives identical physical results to the conditional master equation (\ref{eqn:conBECmasts}). However, it also suffers from the same curse of dimensionality as the conditional master equation. In order to address this issue we look to classical probability. In classical probability theory, all probability distributions whose evolution is continuous in time and Markovian must obey a Fokker-Plank equation \cite{Gardiner:2004}. Fokker-Planck equations suffer from the same curse of dimensionality as quantum systems insofar as the size of the probability distribution grows exponentially with the number of degrees of freedom. Fortuitously, for every Fokker-Planck equation there is an equivalent set of stochastic differential equations (SDEs) that can be simulated efficiently. The moments of the SDEs are simply averaged many times to produce the same expectations as the Fokker-Planck equation. This `trick' is how phase-space methods such as TW escape the curse of dimensionality. Specifically, when applying TW, higher-order derivatives are truncated such that the quasi-probability distribution then obeys a Fokker-Planck equation, which can be unravelled into a set of SDEs. We apply a similar method here with the NPW particle filter. However, since the system is continuously-monitored we instead aim to produce a Kushner-Stratonovich equation which can then be unravelled into a set of WSDEs, as detailed in \cite{Hush:2009}.

We note that the terms in equation~(\ref{eqn:conBECnpw}) generated by the measurement and, perhaps surprisingly, the nonlinear inter-atomic interactions are all in Kushner-Stratonovich form and can be unravelled exactly. However, the terms on the first three lines, which were generated by the single-particle Hamiltonian, are not. Fortunately, in typical BEC experiments, we can safely assume that the number of particles, $N$, is much larger than the number of occupied modes, $M$. This allows us to make the following three approximations:
\begin{enumerate}
\item By definition of the NPW representation, terms of the form $\int d\varphi'(\cdot) e^{2i(\varphi(\mv{y}) - \varphi'(\cdot))(n(\mv{y}) + 1/2)}$ are measures of the state of the system's overlap with the vacuum. If $N \gg M$ then there will only be a negligible occupation of the vacuum, and thus we can simply set these terms to zero.
\item The square roots on the first line of equation~(\ref{eqn:conBECnpw}) can be expanded and truncated to second order in derivatives with respect to $\varphi(\mv{x})$. This is essentially the same approximation as required by TW. The approximation is valid provided functions under the square root are sufficiently smooth, which is guaranteed when $N \gg M$.
\item The discrete variable $n(\cdot)$ is taken to the continuous limit, which is an excellent approximation when $N \gg M$. 
\end{enumerate}
These approximations were all developed and verified in \cite{Hush:2012,Hush:2012a}. After the application of these approximations, equation~(\ref{eqn:conBECnpw}) simplifies to 
\begin{eqnarray}
	\fl d \sN[n(\cdot),\varphi(\cdot)] = \Bigg\{ \Bigg( \frac{1}{\hbar} \int d\mv{x} \int d\mv{y} h'(\mv{x},\mv{y},\mv{u}) %
	\Bigg(  -i \partial_{n(\mv{y})} \sqrt{\left(n(\mv{x}) + \delta(0)/2 \right)\left( n(\mv{y}) + \delta(0)/2 \right) } \nn \\
	\fl \quad + \frac{1}{2} \partial_{\varphi(\mv{y})} \sqrt{\frac{n(\mv{x}) + \delta(0)/2}{n(\mv{y}) + \delta(0)/2}} \Bigg)e^{i(\varphi(\mv{y}) - \varphi(\mv{x}))} + \mbox{c.c.} \Bigg) + U \int d\mv{x} \; \partial_{\varphi(\mv{x})} n(\mv{x}) \nn \\
	\fl \quad + \sum_i \Bigg( \frac{1}{2} \int d\mv{x} \int d\mv{y} \; l_i(\mv{x})l_i(\mv{y}) \partial_{\varphi(\mv{x})} \partial_{\varphi(\mv{y})} - 2 \Big( (L_i^n)^2 - \sE[(L_i^n)^2] - 2 (L_i^n - \sE[L_i^n])\sE[L_i^n]\Big)\nn \\
	\fl \quad  + 2 \left(L_i^n - \sE[L_i^n]\right) \; \eta_i(t) \Bigg) \Bigg\}\sN[n(\cdot),\varphi(\cdot)].\label{eqn:appconBECnpw}
\end{eqnarray}
Equation~(\ref{eqn:appconBECnpw}) is in Kushner-Stratonovich form and can be unravelled into a set of WSDEs. However, it can be expressed in a more convenient form by making the transformation $\alpha(\mv{x}) = \sqrt{n(\mv{x}) + \delta(0)/2}\, e^{i\varphi(\mv{x})}$, where $\alpha(\mv{x})$ is a complex function, before unravelling. Applying this transformation, we find equation~(\ref{eqn:appconBECnpw}) is equivalent to
\begin{eqnarray}
	\fl d\sN[\alpha(\cdot)] = \Bigg\{ \frac{i}{\hbar} \int d\mv{x} \, \Big( \; \partial_{\alpha(\mv{x})} \left(h(\mv{x},\mv{u}) + U |\alpha(\mv{x})|^2 \right) \alpha(\mv{x}) \nn \\
	\qquad - \partial_{\alpha^*(\mv{x})} \left(h(\mv{x},\mv{u}) + U |\alpha(\mv{x})|^2 \right) \alpha^*(\mv{x}) \Big) \nonumber \\
	\fl \, + \sum_i \Bigg( \frac{1}{2} \int d\mv{x} \int d\mv{y} \; l_i(\mv{x})l_i(\mv{y}) \Big(\partial_{\alpha(\mv{x})}\partial_{\alpha(\mv{y})} \alpha(\mv{x}) \alpha(\mv{y}) + 2\partial_{\alpha^*(\mv{x})}\partial_{\alpha(\mv{y})} \alpha^*(\mv{x}) \alpha(\mv{y}) \nn \\
	\qquad + \partial_{\alpha^*(\mv{x})} \partial_{\alpha^*(\mv{y})} \alpha^*(\mv{x}) \alpha^*(\mv{y}) \Big) \nn \\
	\fl \,- 2 \Big( (L_i^\alpha)^2 - \sE[(L_i^\alpha)^2] - 2 (L_i^\alpha - \sE_{\sN}[L_i^\alpha])\sE[L_i^\alpha] \Big) + 2 \left(L_i^\alpha - \sE[L_i^\alpha]\right) \; \eta_i(t) \Bigg) \Bigg\} \sN[\alpha(\cdot)],  \label{eqn:simpconBECnpw}
\end{eqnarray}
where  
\begin{eqnarray}
	L_i^\alpha 	= \int d\mv{x} l_i(\mv{x}) (|\alpha(\mv{x})|^2 - \delta(0)/2), \nn \\
	 \sE\left[f[\alpha(\cdot)]\right]  = \int d\alpha(\cdot) f[\alpha(\cdot)] \sN[\alpha(\cdot)], \label{eqn:appexpectalpha}
\end{eqnarray}
and $d\alpha(\cdot)$ is a functional integration measure. 

Since equation~(\ref{eqn:simpconBECnpw}) is in Kushner-Stratonovich form, it can be unravelled into an equivalent set of WSDEs \cite{Hush:2009}: 
\begin{eqnarray}
	\fl \partial_t \alpha^{(k)}(\mv{x}) = \Big( -\frac{i}{\hbar}(h(\mv{x},\mv{u}) + U|\alpha^{(k)}(\mv{x})|^2)-i \sum_i l_i(x) \zeta_i^{(k)}(t) \Big) \alpha^{(k)}(\mv{x}),  \label{NPW_alpha}\\
	\fl \partial_t w^{(k)} = \Big( \sum_i 2 (2 L_i^{(k)} \mW[L_i^{(\cdot)}] - (L_i^{(k)})^2) + 2 L_i^{(k)} \eta_i(t) \Big)w^{(k)}, \label{NPW_w}
\end{eqnarray}
where $k$ is an integer indexing the $K$ stochastic paths, $\zeta_i^{(k)}(t)$ is a `fictitious' Stratonovich noise increment, which is generated independently for each of the $K$ paths, 
\begin{equation}
	L_i^{(k)} = \int d\mv{x} \; l_i(\mv{x}) \left(|\alpha^{(k)}(\mv{x})|^2 - \frac{1}{2}\delta(x)\right),
\end{equation}
 and 
\begin{equation}
	\mW[f(\alpha^{(\cdot)}(\mv{x}))] \equiv \sum_{k=1}^K \frac{w^{(k)} f(\alpha^{(k)}(\mv{x}))}{w^{(k)}} \label{eqn:appweightedmean}
\end{equation}
is the definition of the weighted average over the swarm. Equations~(\ref{NPW_alpha}) and (\ref{NPW_w}) are restated in the main text as equations~(15) and is what we refer to as the \emph{NPW particle filter}.

Equation~(\ref{eqn:simpconBECnpw}) and the NPW particle filter are related through their weighted averages [Eq~(\ref{eqn:appweightedmean})] and expectations [Eq~(\ref{eqn:appexpectalpha})], respectively. Specifically,  
\begin{equation}
\mW[f[\alpha(\cdot)] ] \equiv \sE[f[\alpha(\cdot)] ], \label{eqn:npqwsdeequiv}
\end{equation}
in the limit that the number of paths, $K$, goes to infinity \cite{Hush:2009}. If a truly infinite number of paths was required for this equality to hold, we would not have solved the curse of dimensionality as claimed. Fortuitously, it has been demonstrated that this equality approximately holds with vastly less memory than would be required to simulate the NPW representation directly. Typically, choosing $K$ somewhere between $10^3-10^6$ works very well for simulations of monitored BECs \cite{Hush:2012,Hush:2012a}. We can use equation~(\ref{eqn:npqmastequiv}) and equation~(\ref{eqn:npqwsdeequiv}) to generate any quantum expectation value from the NPW particle filter. For example, the expectations of the field operator and the one-body correlation correlation matrix are
\begin{eqnarray}
	\langle \hat{\psi}(\mv{x}) \rangle = \mW[ \alpha(\mv{x})], \label{eqn:exampexpectvalues_1} \\
	\langle \hat{\psi}(\mv{x})^\dag \hat{\psi}(\mv{y}) \rangle = \mW[\alpha(\mv{x})^*\alpha(\mv{y})] - \frac{1}{2} \delta(\mv{x}-\mv{y}), \label{eqn:exampexpectvalues_2}
\end{eqnarray}
respectively.

\section{Initial conditions for simulations} \label{apx:sampNPWparfil}

In this appendix we discuss the initial conditions used to simulate equation~(\ref{eqn:conBECmasts}) via both the Hatree-Fock method and the NPW particle filter. To begin, we note that all simulations in this paper assume that the atomic field is in the BEC phase, where all atoms occupy the same quantum state. However, the atoms are \emph{not} initially in the ground state of the trap, and are instead in a far from equilibrium state with an order parameter described by equation~(\ref{eqn:ICdispgauss}). Given that the Hartree-Fock approximation is only truly valid in the zero temperature limit, initializing the field in a BEC state allowed for a fair comparison between semiclassical Hartree-Fock and NPW particle filter simulations. Furthermore, insight into the effectiveness of the feedback control is given by choosing an initial condition with much higher energy than the ground state, as this provides a qualitative picture of the rate at which the feedback removes energy from the system. In principle, this initial condition could be experimentally prepared by lowering interactions in the condensate (through a Feshbach resonance \cite{Leggett:2001}), waiting for the atoms to reach the Gaussian ground state of the trap, and then performing a rapid change in the position of the trap. Of course, this initial condition does not correctly describe finite temperature and strongly-correlated Bose-condensed clouds of atoms. Nevertheless, we anticipate that our conclusions on the effectiveness of the feedback control qualitatively hold more generally, as the stability and robustness of the feedback mechanism ensures that the long-term behaviour of the system is independent of the initial condition.

Although the Hartree-Fock method and the NPW particle filter use the same initial condition [equation~(\ref{eqn:ICdispgauss})], the numerical implementation is very different. The Hartree-Fock method simply sets the value of the order parameter $\phi(\mv{x},t)$ at $t=0$. In contrast, the NPW particle filter requires a sampling of the swarm of fields, $\alpha^{(k)}(\mv{x})$, and weights, $w^{(k)}$, from an appropriate NPW representation of the initial state's density matrix. The remainder of this appendix outlines the details of these implementations.

For a BEC at temperature $T \approx 0$\,K, which is well-described by the mean-field wavefunction $\alpha_0(\mv{x})$ \cite{Dalfovo:1999}, the density matrix is 
\begin{equation}
	\hat{\rho}_{\alpha_0} = \int_0^{2\pi} d\theta \, |\alpha_0(\cdot) e^{i\theta} \tensor*{\rangle}{_c} \tensor*[_c]{\langle}{} \alpha_0(\cdot) e^{i\theta} |, \label{eqn:denscohstate}
\end{equation}
where $ |\alpha_0(\cdot)\rangle_c$ means a field of coherent states that has the property
\begin{equation}
 	\hat{\psi}(\mv{x})|\alpha_0(\cdot)\rangle_c = \alpha_0(\mv{x}) |\alpha_0(\cdot)\rangle_c.
\end{equation}
The integral over $\theta$ is included because the mean field of a BEC is known only up to an absolute phase. This density matrix can equivalently be written (in first quantized notation) as a mixture of the Hartree-Fock states presented in equation~(\ref{eqn:denssamewave}):
\begin{equation}
	\hat{\rho}_{\alpha_0} = \sum_{n=0}^{\infty} \frac{N^n e^{-N}}{n!} \bigotimes_{i=1}^n \left(\int d\mv{x} \, d\mv{y} \, \tilde{\alpha}_0(\mv{x},t) \tilde{\alpha}_0^*(\mv{y},t) |\mv{x}  \tensor*{\rangle}{_i} \tensor*[_i]{\langle}{} \mv{y} |\right), \label{eqn:denscohstateprime}
\end{equation}
where $N = \int d\mv{x} |\alpha_0(\mv{x})|^2$ and $\tilde{\alpha}_0(\mv{x}) = \alpha_0(\mv{x})/\sqrt{N}$. Note that the probability distribution for the total number of atoms in the condensate is a Poissonian distribution with mean $N$. The equivalence of representations (\ref{eqn:denscohstate}) and (\ref{eqn:denscohstateprime}) reaffirms that the Hartree-Fock approximation and coherent-state-based mean-field approximation are closely related. 

Since the Hartree-Fock approximation assumes that the total number of atoms in the condensate is \emph{fixed}, the representation (\ref{eqn:denscohstateprime}) cannot strictly be used as the initial state in Hartree-Fock simulations. However, if $N \gg 1$ then the uncertainty in the total number of atoms is small. For Hartree-Fock simulations, we are therefore justified in setting the total atom number to the mean number $N$ and choosing the initial condition 
\begin{equation}
	\phi(\mv{x},t=0) = \frac{\alpha_0(\mv{x})}{\sqrt{N}}.
\end{equation}

For the NPW particle filter, the initial conditions must be chosen such that weighted averages give the same values as the corresponding quantum expectation values generated from equation~(\ref{eqn:denscohstate}). As with phase-space methods such as TW, the obvious distribution from which to sample from is the NPW quasi-probability representation for the initial density matrix (\ref{eqn:denscohstate}):
\begin{equation}
\sN_{\alpha_0}[n(\cdot),\phi(\cdot)] = \sum_{k(\cdot)=-n(\cdot)}^{n(\cdot)} \prod_\mv{x} \frac{n_0(\mv{x})^{n(\mv{x})} e^{2ik(\phi(\mv{x}) - \phi_0(\mv{x})) - n_0(\mv{x})}}{2\pi\sqrt{(n(\mv{x})-k(\mv{x})))!(n(\mv{x})+k(\mv{x}))!}}, \label{eqn:truenpqcohstate}
\end{equation}
where $n_0(\mv{x}) = |\alpha_0(\mv{x})|^2$ and $\varphi_0(\mv{x}) = \mbox{arg}[\alpha_0(\mv{x})]$. However, the quasi-probability distribution in equation~(\ref{eqn:truenpqcohstate}) can have negative values, and thus cannot be efficiently sampled \cite{Troyer:2005}. Instead, we sample from an \emph{approximate} distribution that is valid when the number of particles $N$ is much larger than the number of simulated modes $M$. The details of these approximations and a derivation of the resulting approximate distribution can be found in \cite{Hush:2012,Hush:2012a}; here we simply present the result: 
\begin{equation}
\sN_{\alpha_0}[n(\cdot),\varphi(\cdot)] \approx \prod_\mv{x} \frac{ \sqrt{2} \, n_0(\mv{x})^{n(\mv{x})} e^{-n_0(\mv{x}) + \frac{-2((\varphi(\mv{x}) - \varphi_0(\mv{x})) }{ \psi^{(1)}(n(\mv{x})+1) }}}{n(\mv{x})! \sqrt{\pi \psi^{(1)}(n(\mv{x})+1)}}, \label{eqn:npqcohstate}
\end{equation}
where $\psi^{(1)}(x)$ is the trigamma function. In essence, this probability distribution is the product of a Poissonian distribution for the number variable $n(\mv{x})$ multiplied by a Gaussian distribution for the phase variable $\varphi(\mv{x})$, for every point in space. When sampling this distribution we keep $n(\mv{x})$ quantized, even though it is treated as a continuous variable during evolution. Importantly, this sampling combined with the evolution of the NPW particle filter [equations~(\ref{NPW_alpha}) and (\ref{NPW_w})] ensures that the \emph{total} number of atoms per sample remains quantized. 

Sampling from equation~(\ref{eqn:npqcohstate}) only approximately generates quantum moments equivalent to equation~(\ref{eqn:truenpqcohstate}). On the occasions when sampling returns $n(\mv{x}) = 0$, the moments generated become much less accurate. This is because the Gaussian distribution for the phase variable is not flat, which means that there will be some phase information in the sample of $\varphi(\mv{x})=0$ when $n(\mv{x})=0$. In other words, sampling ascribes some phase information when we occasionally sample vacuum states, which is not physically appropriate. To remedy this situation, we treat times when we sample $n(\mv{x})=0$ as special cases where $\varphi(\mv{x})$ is instead sampled from a flat distribution across phase rather than from a Gaussian distribution. As shown more completely in \cite{Hush:2012a}, the application of this technique tends to vastly improve the accuracy of the quantum moments generated. Even with this improvement the sampling technique still only valid $N \gg M$ (meaning the number of particles $N$ is much larger than the number of occupied modes $M$); the same approximate limit the evolution is valid in.

To summarize, we present the following algorithm for sampling the swarm that makes up a NPW particle filter when initial state is a BEC with mean field $\alpha_0(\mv{x})$: 
\begin{enumerate}
\item Run steps (ii) - (vi) for every position $\mv{x}$ and $k \in  (1, K)$.
\item Sample a random variable from a Poisson distribution with mean $n_0(\mv{x}) = |\alpha_0(\mv{x})|^2/\delta(0)$, and store it in $n^{(k)}$. 
\item If $n^{(k)}=0$ go to (\ref{lst:vacuumyes}), else go to (\ref{lst:vacuumno}).
\item \label{lst:vacuumyes} Sample a random variable from a uniform distribution over the interval $[0,2\pi)$, and store it in $\varphi^{(k)}(\mv{x})$. Go to (vi).
\item \label{lst:vacuumno} Sample a random variable from a Gaussian distribution with mean $\varphi_0(\mv{x}) = \mbox{arg}[\alpha_0(\mv{x})]$ and variance $\frac{1}{4}\psi^{(1)}(n(\mv{x})+1)$, store it in $\varphi^{(k)}(\mv{x})$.
\item Store the weight $w^{(k)} = 1$ and stochastic field 
	\begin{equation}
		\alpha^{(k)}(\mv{x}) = \sqrt{(n+ 1/2)\delta(0)}e^{i\varphi(\mv{x})}.
	\end{equation}
\end{enumerate}
Once the swarm has been sampled it can the be integrated using equations~(\ref{NPW_alpha}) and (\ref{NPW_w}). We must perform a new and independent sample of the swarm every time we integrate a new virtual measurement record, $\eta_i(t)$, or, in other words, each time we perform a new `experiment'.

Finally, we note that the initial state sampling and the validity of the simulations are intimately related. As previously stated, simulations must have a large particle number $N$ compared with the number of occupied modes $M$ to ensure that both the initial sampling and evolution of the NPW particle filter stay accurate. For all simulations performed in this paper, the number of particles was $N=100$ and the number of occupied modes was less than $M \approx N/3$, which satisfies the approximation. We were unable to perform simulations with a higher number of atoms while the system was being monitored. Simulating a BEC under measurement involves a process of evolving and resampling sets of weights, which becomes more numerically demanding as the number of atoms increase. Fortuitously, the NPW particle filter is \emph{exact} for the terms corresponding to the measurement, harmonic trapping potential and the nonlinear inter-atomic interactions. Consequently, the requirement that $N \gg M$ is entirely due to the approximate form of the kinetic energy term. Thus, the validity of the approximations on the kinetic energy term can be verified \emph{independent} of the measurement. In particular, simulations were performed on a BEC without an inter-atomic nonlinearity or measurement using the NPW particle filter and compared to an exact solution. They were found to match closely in the $N=100$ particle limit considered in this paper. 

\section*{References}
\bibliographystyle{unsrt.bst}
\bibliography{cancelquantnoise_papers}

\end{document}